\documentclass[aps,prresearch,preprint,superscriptaddress,amsmath,amssymb,floatfix]{revtex4-2}

\usepackage{graphicx}
\usepackage{bm}
\usepackage{braket}
\usepackage{hyperref}
\usepackage{xcolor}

\begin{document}

\title{Transfer tensor description of driven inhomogeneous quantum systems}

\author{Ferdos Fessahaye}
\affiliation{Department of Chemistry and Biochemistry, University of Nevada, Las Vegas, Nevada 89154, USA}

\author{Rajesh Dutta}
\affiliation{Department of Chemistry and Biochemistry, University of Nevada, Las Vegas, Nevada 89154, USA}

\author{Michelle Anderson}
\affiliation{Department of Chemistry and Biochemistry, University of Nevada, Las Vegas, Nevada 89154, USA}

\author{Chern Chuang}
\email{chern.chuang@unlv.edu}
\affiliation{Department of Chemistry and Biochemistry, University of Nevada, Las Vegas, Nevada 89154, USA}

\date{\today}

\begin{abstract}
The transfer tensor method (TTM) provides an efficient, non-perturbative framework to extrapolate open quantum system dynamics from short-time trajectories by encoding memory effects into a discrete hierarchy of transfer tensors. 
Standard formulations rely on time-translational invariance (TTI) of the dynamical maps, which is explicitly broken when a system is subjected to time-dependent driving or coherent control fields. In this work, we generalize TTM to time-inhomogeneous and periodically driven quantum systems, where a two-time transfer tensor hierarchy $\mathcal{T}(n,k)$ propagates reduced density matrices across arbitrary time steps. For time-periodic driving with period $t_0$, sub-stroboscopic sampling ($\Delta t = t_0 / N_c$) restores a discrete block TTI, $\mathcal{T}(n+mN_c, k+mN_c) = \mathcal{T}(n,k)$, so that multi-cycle propagation requires learning only $N_c$ starting-time branches. We further introduce temporal homogenization by averaging over the driving phase (equivalently, the start time), which restores single-index TTI at the price of phase-induced memory that combines non-additively with that of static disorder. We demonstrate the framework on two paradigmatic systems: (i) an inhomogeneously broadened two-level ensemble driven beyond the rotating-wave approximation (RWA), where the phase-induced memory scales with the square of the Rabi coupling and the disorder-induced memory with the square of the inhomogeneous width, so that the two become comparable when these energy scales are. This phase-induced memory depends on the reference frame: in the frame rotating with the drive it arises solely from counter-rotating terms and vanishes under the RWA, whereas in the lab frame it largely survives the RWA; and (ii) a driven tight-binding chain at the Dunlap--Kenkre dynamical localization condition. Stroboscopically, the clean chain is exactly frozen and carries neither memory nor pseudoflux, whereas sub-stroboscopic transfer tensors reveal that dynamical localization is enforced by delayed, period-commensurate refocusing in the memory kernel, in contrast to the immediate backflow characteristic of Anderson localization. A generalized outgoing pseudoflux tracks the non-monotonic dependence of the spreading on disorder strength: moderate disorder breaks the refocusing and lets the particle spread over a larger, but still finite, region, whereas strong disorder confines it by Anderson localization.
\end{abstract}

\maketitle

\section{Introduction}
\label{sec:intro}
Describing the non-equilibrium quantum dynamics of open systems subject to both environmental dissipation and external driving is a central challenge across coherent spectroscopy, quantum transport, and quantum information processing~\cite{Breuer2002,Follia2026}. When a quantum system interacts with a structured reservoir possessing finite correlation times, or when an ensemble of emitters exhibits static structural or energetic inhomogeneity, the reduced dynamics departs strongly from a memoryless dynamical semigroup. Among non-perturbative numerical approaches to non-Markovian open-system dynamics, the transfer tensor method (TTM) introduced by Cerrillo and Cao~\cite{Cerrillo2014} offers a particularly transparent and data-efficient framework. By recursively decomposing a family of short-time dynamical maps $\mathcal{M}(n)$ into discrete transfer tensors $\mathcal{T}(k)$ that play a role similar to a time-discretized Nakajima--Zwanzig memory kernel~\cite{Cerrillo2014,MorillasRozas2026}, TTM captures environmental memory up to a cutoff depth $K$ and extrapolates the reduced density matrix to arbitrarily long times. TTM has been combined with numerically exact propagators including hierarchical equations of motion, path integrals, mixed quantum-classical Liouville dynamics~\cite{Kananenka2016}, and stochastic Schr{\"o}dinger equations~\cite{Gao2026}, and extended to accommodate initial system--bath correlations~\cite{Buser2017} and multi-time correlation functions~\cite{Jorgensen2020,Gherardini2022}. 
The applicability of the TTM was examined in different regimes of system-bath parameterizations, and it was demonstrated that the TTM provides an efficient approach to accurately simulate long-time open quantum system dynamics, particularly when the applicability of perturbative approaches such as F\"orster and Redfield theories is uncertain~\cite{gelzinis2017applicability}. Its first-principles derivation from the process-tensor framework shows that transfer tensors can be reconstructed from system-only process tomography even for driven or initially correlated systems~\cite{PollockModi2018}. 
A TTM based approach was used to characterize non-Markovian quantum dynamics and reconstruct the memory kernel governing the interaction between a quantum system and its environment using quantum process tomography~\cite{chen2020non}. The approach further enables the extraction of environmental noise correlation functions and spectral densities, and was experimentally demonstrated on an IBM quantum processor through the identification of non-Markovian dissipation and spatially correlated noise. Recently, TTM has been employed alongside the generalized quantum master equation to estimate memory times and extend nonadiabatic dynamics from short-time simulations to long times, with the accuracy of both approaches benchmarked using spin--boson and multistate harmonic models, including a realistic carotenoid--porphyrin--fullerene triad~\cite{zeng2026estimating}. A recent analytical study clarifies the distinction between the TTM and the Nakajima--Zwanzig memory kernel at finite time steps and examines the emergence of effectively Markovian behavior in a non-Markovian Jaynes--Cummings system under specific time discretizations~\cite{MorillasRozas2026}.

Despite these advances, conventional TTM rests on the assumption of time-translational invariance (TTI): the underlying Liouvillian and memory kernel are assumed stationary so that the two-time dynamical map $\mathcal{M}(t_n, t_k)$ depends solely on the elapsed duration $t_n - t_k$. However, across a wide variety of experimental settings---ranging from room-temperature optical-cycling of molecular qubits in liquid cells~\cite{Shin2024} and semiconductor quantum dots to Floquet-engineered cold-atom and photonic lattices---the system Hamiltonian $H(t) = H_0 + V(t)$ is explicitly time-dependent. Periodic driving can reshape steady-state coherence and energy transport in open systems~\cite{Cao2023,Cao2026} and can be used to control memory effects~\cite{Follia2026}, while the commonly used Floquet master equations are restricted to weak or high-frequency driving~\cite{Mickiewicz2026}. Driven spin--boson dynamics have also been described using a Bloch--Redfield master equation, where time-dependent control fields modify the dissipative rates and are optimized to implement high-fidelity single-qubit gates~\cite{jirari2026analytical}. Numerically exact descriptions of strongly damped, periodically driven systems have recently been formulated with periodic influence-functional (matrix-product) propagators~\cite{Mickiewicz2025}. 
Continuous or pulsed driving breaks TTI, rendering standard single-index transfer tensors inapplicable whenever sub-cycle resolution is required.

At the same time, recent work by Anderson and Chuang~\cite{Anderson2026} demonstrated that even when individual trajectory evolutions are Markovian, ensemble averaging over static spatial or energetic disorder breaks the dynamical semigroup property due to the non-convex geometry of Markovian dynamical maps. Denoting ensemble averages with a bar for an ensemble of $N_{\text{ens}}$ trajectories,
\begin{equation}
\begin{aligned}
\bar{\mathcal{T}}(2) &= \bar{\mathcal{M}}(2) - [\bar{\mathcal{M}}(1)]^2 \\
&= \frac{1}{N_{\text{ens}}}\sum_{i=1}^{N_{\text{ens}}} [\mathcal{M}^{(i)}(1)]^2 - \left[\frac{1}{N_{\text{ens}}}\sum_{i=1}^{N_{\text{ens}}} \mathcal{M}^{(i)}(1)\right]^2 \neq 0.
\end{aligned}
\label{eq:spatial_memory_intro}
\end{equation}
Physically, this reflects the principle that \emph{resolution loss induces memory}: once fine-grained specification of which static disorder realization a trajectory belongs to is discarded by ensemble averaging, the coarse-grained dynamics must retain memory to distinguish static disorder from uncorrelated dynamical noise~\cite{Anderson2026}.
Memory generated by mixing Markovian evolutions is well documented~\cite{Breuer2018,Lorenzo2017} and is classified as classical, or non-genuine, non-Markovianity~\cite{Gangwar2026}; disorder averaging can even suppress the observable signatures of memory while the averaged dynamics remains indivisible~\cite{Sengupta2026}. Throughout this work, \emph{memory} refers to nonzero transfer tensors beyond a single step, $\mathcal{T}(n,k) \neq 0$ for $n-k \ge 2$, i.e., to the non-divisibility of the averaged maps~\cite{Gherardini2022}; we do not relate the tensor norms to quantitative measures of non-Markovianity. 
Other criteria can classify the same ensemble differently: trace-distance and decay-rate witnesses find Gaussian-disorder-averaged qudit dynamics Markovian~\cite{Santra2024}, and the disorder-averaged single-particle self-energy entering nonequilibrium Green's functions can be time-local~\cite{Chaudhary2026}. A classical counterpart of resolution-loss-induced memory appears in coarse-grained Langevin dynamics, where a mean-field treatment absorbs static spatial heterogeneity into an inflated memory kernel~\cite{Liu2026}. Static inhomogeneity also governs transport in condensed-phase settings such as exciton-polaritons in disordered microcavities~\cite{Engelhardt2023,Aroeira2025}.

In this paper, we extend both the mathematical machinery of TTM and the principle of coarse-graining-induced memory to time-inhomogeneous and periodically driven quantum systems. First, we formulate a general two-index transfer tensor hierarchy $\mathcal{T}(n,k)$ for arbitrary time-dependent Hamiltonians $H(t)$. Second, for periodically driven systems $V(t + t_0) = V(t)$ sampled sub-stroboscopically ($\Delta t = t_0 / N_c$ with integer $N_c > 1$), we show that the two-index tensors obey a discrete block periodicity, $\mathcal{T}(n + mN_c, k + mN_c) = \mathcal{T}(n,k)$, which restricts the learning requirement to $N_c$ starting times within a single driving period while retaining full intra-cycle resolution. Third, we introduce \emph{temporal homogenization} by averaging over the initial phase $\phi$ of the periodic drive (or equivalently over starting times within a period). This procedure restores TTI and reduces the tensors to single indexing at the cost of generating \emph{phase-induced memory}, which combines non-additively with static energetic disorder.

We illustrate the framework on two benchmark problems. First, we study an inhomogeneously broadened ensemble of two-level systems driven by a monochromatic field~\cite{Noh2010,Shin2024}. Averaging over the driving phase generates memory that grows quadratically with the Rabi coupling, whereas the memory generated by the static inhomogeneity grows quadratically with the inhomogeneous width; their relative size is set by the ratio of the two, with a crossover when the inhomogeneous width becomes comparable to the Rabi coupling. 
 Because the start-time average does not commute with a time-dependent change of frame, this memory depends on the frame in which the averaged dynamics is described: in the frame rotating with the drive it is caused entirely by counter-rotating terms and is erased by the rotating-wave approximation (RWA), whereas in the lab frame the RWA retains most of it. 
Second, we investigate a 1D tight-binding chain driven at the Dunlap--Kenkre dynamical localization condition~\cite{Dunlap1986} in the presence of static Anderson disorder~\cite{Anderson1958}. Extending the outgoing pseudoflux of Ref~\cite{Anderson2026}. to driven systems, we show that stroboscopic sampling sees the clean dynamically localized chain as trivially frozen, whereas sub-stroboscopic transfer tensors expose the mechanism of dynamical localization---delayed refocusing at integer multiples of the driving period---and distinguish it sharply from the immediate backflow underlying Anderson localization. The pseudoflux also tracks the non-monotonic dependence of the spreading on disorder: intermediate disorder breaks the refocusing and enlarges the region over which the particle spreads, without delocalizing it, whereas strong disorder confines it by Anderson localization.

The remainder of this paper is organized as follows. Section~\ref{sec:theory} develops the two-time transfer tensor formalism, sub-stroboscopic block invariance, temporal homogenization, and the generalized outgoing pseudoflux. Section~\ref{sec:examples} presents numerical results for the driven two-level ensemble and the driven disordered tight-binding chain. Section~\ref{sec:conclusion} summarizes our conclusions and outlook. Appendix~\ref{app:column} shows how the tensor quantities used for the lattice are evaluated without storing full superoperators, Appendix~\ref{app:numerics} collects numerical details, and Appendix~\ref{app:scaling} derives how the magnitude of averaging-induced memory scales with the inhomogeneous width, the drive amplitude and waveform, and the time step. 

\section{Two-Time Transfer Tensor Formalism}
\label{sec:theory}

\subsection{\texorpdfstring{Arbitrary Time-Dependent Driving and Two-Index Tensors $\mathcal{T}(n,k)$}{Arbitrary Time-Dependent Driving and Two-Index Tensors T(n,k)}}
\label{subsec:theory_twoindex}
Consider an open quantum system with reduced density matrix $\rho(t) = \mathrm{Tr}_{\text{B}}[\rho_{\text{tot}}(t)]$ evolving under a time-dependent Hamiltonian $H(t) = H_0 + V(t)$. Assuming a factorized initial state at preparation time $t_k = k\Delta t$, $\rho_{\text{tot}}(t_k) = \rho(t_k) \otimes \rho_{\text{B}}^{\text{eq}}$, the evolution from step $k$ to step $n \ge k$ on a uniform time grid $t_n = n\Delta t$ is governed by a linear two-time dynamical map $\mathcal{M}(n,k)$:
\begin{equation}
\rho(n) = \mathcal{M}(n,k) \rho(k), \quad \text{with} \quad \mathcal{M}(k,k) = \mathbb{I}.
\label{eq:dyn_map_def}
\end{equation}
When $V(t) \neq 0$, TTI is broken, so $\mathcal{M}(n,k)$ depends separately on both the final time index $n$ and the starting time index $k$, forming a lower-triangular hierarchy of superoperators indexed by $(n,k)$ with $n \ge k$.

To isolate the direct dynamical correlation between step $k$ and step $n$ from intermediate multi-step pathways, a set of two-index transfer tensors $\mathcal{T}(n,k)$ ($n > k$) can be recursively defined by the convolution structure of the dynamical maps:
\begin{equation}
\mathcal{T}(n, k) = \mathcal{M}(n, k) - \sum_{l=k+1}^{n-1} \mathcal{T}(n, l) \mathcal{M}(l, k),
\label{eq:tt_two_index_recursion}
\end{equation}
initialized by the single-step propagators along the sub-diagonal,
\begin{equation}
\mathcal{T}(n, n-1) = \mathcal{M}(n, n-1).
\label{eq:tt_Explicit_base}
\end{equation}
Unrolling Eq.~\eqref{eq:tt_two_index_recursion} for initial step $k=0$ yields explicitly:
\begin{align}
\mathcal{T}(1,0) &= \mathcal{M}(1,0), \label{eq:tt_explicit_1} \\
\mathcal{T}(2,0) &= \mathcal{M}(2,0) - \mathcal{T}(2,1) \mathcal{M}(1,0), \label{eq:tt_explicit_2} \\
\mathcal{T}(3,0) &= \mathcal{M}(3,0) - \mathcal{T}(3,1) \mathcal{M}(1,0) - \mathcal{T}(3,2) \mathcal{M}(2,0). \label{eq:tt_explicit_3}
\end{align}
For a time-local (Markovian) evolution where the two-time maps satisfy the divisible composition law $\mathcal{M}(n,k) = \mathcal{M}(n,n-1)\mathcal{M}(n-1,n-2)\cdots\mathcal{M}(k+1,k)$, substitution into Eq.~\eqref{eq:tt_two_index_recursion} gives $\mathcal{T}(n,k) = 0$ identically for all $n - k \ge 2$, regardless of how strongly $H(t)$ varies with time. Nonzero higher-order tensors $\mathcal{T}(n,k)$ with $n - k \ge 2$ therefore signal memory, i.e., non-divisibility of the maps, across the interval\ 
 $[t_k, t_n]$.

By combining Eqs.~\eqref{eq:dyn_map_def} and \eqref{eq:tt_two_index_recursion}, the reduced density matrix prepared at $k_0 = 0$ is propagated to step $n$ via the closed two-time convolution:
\begin{equation}
\rho(n) = \sum_{k=\max(0, n-K)}^{n-1} \mathcal{T}(n, k) \rho(k),
\label{eq:tt_propagation}
\end{equation}
where $K$ is the memory truncation depth beyond which the tensor's Frobenius norm $\|\mathcal{T}(n, n-m)\|_F$ falls below a specified accuracy threshold.

For an inhomogeneous ensemble of $N_{\text{ens}}$ independent systems characterized by static parameters $\{\xi^{(i)}\}$ (such as static site-energy disorder or inhomogeneous optical detunings), the ensemble-averaged reduced density matrix $\bar{\rho}(n) = N_{\text{ens}}^{-1}\sum_i \rho^{(i)}(n)$ evolves under the ensemble-averaged two-time dynamical map:
\begin{equation}
\bar{\mathcal{M}}(n,k) = \frac{1}{N_{\text{ens}}} \sum_{i=1}^{N_{\text{ens}}} \mathcal{M}^{(i)}(n,k).
\label{eq:ensemble_map}
\end{equation}
(For a continuous distribution of static parameters, the sum is replaced by the corresponding weighted average or quadrature.) Replacing $\mathcal{M}(n,k)$ by $\bar{\mathcal{M}}(n,k)$ in Eq.~\eqref{eq:tt_two_index_recursion} defines the ensemble-averaged two-index transfer tensors $\mathcal{T}(n,k)$. Even when each individual realization $i$ evolves under a Markovian time-dependent Lindblad equation so that $\mathcal{M}^{(i)}(n,k) = \mathcal{M}^{(i)}(n,n-1)\cdots\mathcal{M}^{(i)}(k+1,k)$, the ensemble average breaks divisibility:
\begin{equation}
\begin{aligned}
\mathcal{T}(k+2, k) &= \frac{1}{N_{\text{ens}}}\sum_{i=1}^{N_{\text{ens}}} \mathcal{M}^{(i)}(k+2,k+1)\mathcal{M}^{(i)}(k+1,k) \\
&\quad - \bar{\mathcal{M}}(k+2,k+1)\bar{\mathcal{M}}(k+1,k) \neq 0,
\end{aligned}
\label{eq:two_time_cov_memory}
\end{equation}
which equals the ensemble covariance of consecutive single-step propagators. Appendix~\ref{app:scaling} extends this covariance form to all lags and derives from it the scaling of the memory magnitude with the averaged parameters. 
The two-time hierarchy therefore separates two distinct sources of time dependence. The first is the explicit dependence of the individual dynamical maps on the preparation time, which arises from the time-dependent driving. The second is the memory generated by averaging over different realizations, even when each realization is individually divisible. These effects are conceptually distinct: the two-time formulation retains the temporal dependence of the underlying dynamics, whereas the higher-order transfer tensors quantify the non-divisibility of the averaged maps.

\subsection{Periodically Driven Systems and Sub-Stroboscopic Symmetry}
\label{subsec:theory_periodic}
For a general aperiodic drive $V(t)$, constructing $\mathcal{T}(n,k)$ requires learning a new column of maps for every starting time step $k$. Furthermore, propagating $n-k$ steps down the column requires all tensors covering the whole triangular area enclosed within the points $(k,k)$, $(n,k)$, and $(n,n)$. This offers no asymptotic compression. A crucial simplification occurs when the driving field is periodic with period $t_0 = 2\pi/\omega$:
\begin{equation}
V(t + t_0) = V(t).
\label{eq:periodic_H}
\end{equation}
Choosing the discretization step $\Delta t = t_0 / N_c$ such that $N_c \in \mathbb{Z}_{\ge 1}$ sub-steps span a single driving period, the two-time dynamical maps inherit discrete Floquet shift invariance under translation of both time arguments by any integer multiple $m \in \mathbb{Z}_{\ge 0}$ of $N_c$:
\begin{equation}
\bar{\mathcal{M}}(n + m N_c, k + m N_c) = \bar{\mathcal{M}}(n, k).
\label{eq:map_periodicity}
\end{equation}

In the \emph{stroboscopic limit} ($N_c = 1$, $\Delta t = t_0$), the system is observed only at integer multiples of the driving period. Equation~\eqref{eq:map_periodicity} then reduces to $\bar{\mathcal{M}}(n,k) = \bar{\mathcal{M}}(n-k, 0)$, recovering single-index TTI, $\mathcal{T}(n,k) = \mathcal{T}(n-k)$. While stroboscopic TTM is computationally minimal, it is blind to intra-cycle dynamics, including sub-cycle Rabi flopping, counter-rotating micromotion, and the intra-cycle excursions of dynamically localized particles discussed in Sec.~\ref{subsec:results_substrobo}.

To resolve intra-cycle dynamics, we choose $N_c > 1$ (\emph{sub-stroboscopic sampling}). By induction on the lag $d = n - k \ge 1$ in Eq.~\eqref{eq:tt_two_index_recursion}, the periodicity of $\bar{\mathcal{M}}$ in Eq.~\eqref{eq:map_periodicity} directly implies \emph{block time-translational invariance} of the two-index transfer tensors [illustrated schematically in Fig.~\ref{fig:schematic}(a)]:
\begin{equation}
\mathcal{T}(n + m N_c, k + m N_c) = \mathcal{T}(n, k), \quad \forall m \in \mathbb{Z}_{\ge 0}.
\label{eq:tt_block_symmetry}
\end{equation}
Consequently, any tensor $\mathcal{T}(n,k)$ is uniquely determined by the memory lag $d = n - k \ge 1$ and the starting-time residue $\kappa = k \bmod N_c \in \{0, 1, \dots, N_c - 1\}$:
\begin{equation}
\mathcal{T}(n, k) = \mathcal{T}(\kappa + d, \kappa), \quad (\kappa = k \bmod N_c,\; d = n - k).
\label{eq:tt_modulo_form}
\end{equation}
Thus, rather than computing an ever-growing triangular array of two-time maps, one only needs to learn $N_c$ columns of dynamical maps $\bar{\mathcal{M}}(\kappa + d, \kappa)$ for $\kappa \in \{0, \dots, N_c - 1\}$ up to the memory cutoff depth $d \le K$ (i.e., a parallelogram of side lengths $K$ and width $N_c$). Equation~\eqref{eq:tt_propagation} can then propagate the sub-stroboscopic state $\rho(n)$ to arbitrarily many driving cycles.

\subsection{Temporal Homogenization via Driving-Phase Averaging}
\label{subsec:theory_phaseavg}
Block time-translational invariance reduces the number of independent starting times, but it does not eliminate the dependence on the starting position within a driving period. A single-index description can be recovered by averaging over these starting positions, or equivalently over the initial phase of the periodic drive. We refer to this operation as temporal homogenization. It restores time-translational invariance at the level of the averaged maps, but the resulting maps generally become non-divisible. Thus, the reduction from a two-index to a single-index hierarchy is accompanied by the generation of phase-induced memory.

\begin{figure}[htbp]
\centering
\includegraphics[width=0.92\textwidth]{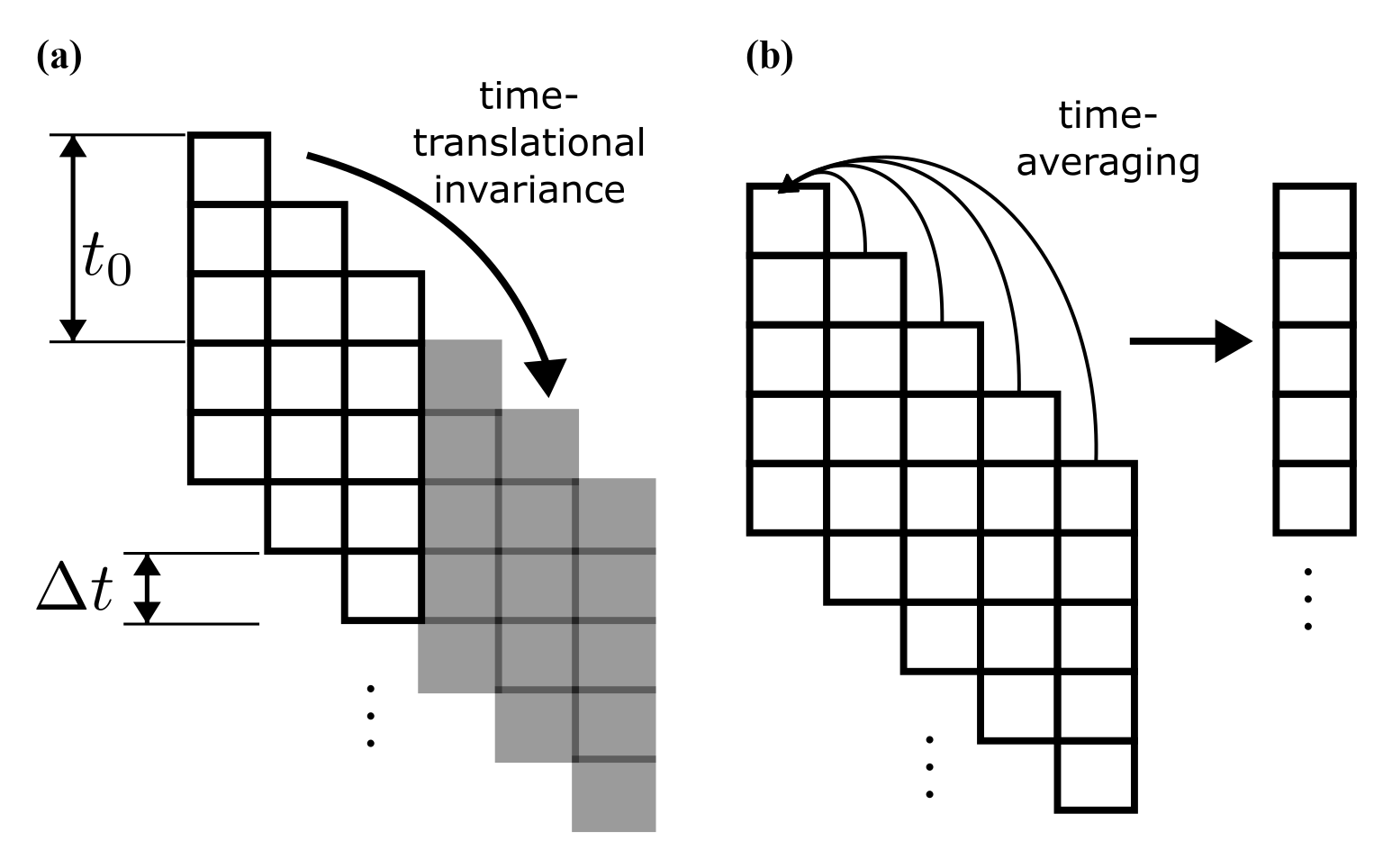}
\caption{Schematic representation of the two-time transfer tensor formalism for periodically driven systems. (a) Lower-triangular dynamical map hierarchy $\mathcal{M}(n,k)$ exhibiting block time-translational invariance under sub-stroboscopic sampling ($\Delta t = t_0 / N_c$). The learning requirement is compressed to $N_c$ columns within a single driving period. (b) Temporal homogenization via start-time (or driving-phase) averaging over a driving period, which collapses the two-index hierarchy into a single-index time-translationally invariant memory series $\bar{\mathcal{T}}_{\text{t-avg}}(d)$.}
\label{fig:schematic}
\end{figure}

This averaging is also motivated by typical experimental realizations of driven ensembles, in which the initial phase $\phi\in[0,2\pi)$ of the periodic field $V(t)=V\cos(\omega t+\phi)$ may vary across individual emitters or experimental shots. Such phase variations arise naturally, for example, when emitters are distributed over a spatial volume larger than the driving wavelength, so that $\phi_i = \mathbf{k}_d \cdot \mathbf{r}_i$ differs between emitters, or in multi-shot pump-probe measurements lacking carrier-envelope phase locking. Because shifting the initial driving phase by $\Delta \phi = 2\pi j / N_c$ is equivalent to shifting the starting time by $j\Delta t$, averaging over the $N_c$ sub-stroboscopic starting times within a period defines the \emph{temporally homogenized} dynamical map [Fig.~\ref{fig:schematic}(b)]:
\begin{equation}
\bar{\mathcal{M}}_{\text{t-avg}}(d) = \frac{1}{N_c} \sum_{\kappa=0}^{N_c - 1} \bar{\mathcal{M}}(\kappa + d, \kappa) \xrightarrow{N_c \to \infty} \frac{1}{2\pi}\int_0^{2\pi} d\phi\, \bar{\mathcal{M}}_\phi(t_d, 0).
\label{eq:phase_avg_map}
\end{equation}
Start-time averaging thus samples the driving phase on the grid $\phi = 2\pi\kappa/N_c$. In the stroboscopic limit $N_c = 1$ the start-time average is trivial, and phase averaging must instead be carried out explicitly over a grid of initial phases $\phi$, as done in Sec.~\ref{subsec:results_dk}.

Temporal homogenization collapses the $N_c$ columns of the two-index hierarchy into a single, time-translationally invariant one-index transfer tensor series:
\begin{equation}
\bar{\mathcal{T}}_{\text{t-avg}}(d) = \bar{\mathcal{M}}_{\text{t-avg}}(d) - \sum_{m=1}^{d-1} \bar{\mathcal{T}}_{\text{t-avg}}(d - m) \bar{\mathcal{M}}_{\text{t-avg}}(m).
\label{eq:tt_tavg}
\end{equation}
The temporally homogenized tensors propagate the phase-averaged density matrix:
\begin{equation}
\bar{\rho}_{\text{t-avg}}(n) = \sum_{d=1}^{\min(K,n)} \bar{\mathcal{T}}_{\text{t-avg}}(d)\,\bar{\rho}_{\text{t-avg}}(n-d),
\label{eq:tavg_propagation}
\end{equation}
capturing the cycle-averaged envelope while filtering out phase-dependent sub-cycle micromotion.

Crucially, temporal homogenization is a coarse-graining operation in the time domain and therefore generates its own \emph{phase-induced memory}. Even for a single homogeneous emitter ($\sigma = 0$) whose fixed-phase evolution is strictly Markovian ($\mathcal{T}(n,k) = 0$ for $n - k \ge 2$), the two-step temporally averaged tensor is given by the variance across starting phases:
\begin{equation}
\begin{aligned}
\bar{\mathcal{T}}_{\text{t-avg}}(2) &= \frac{1}{N_c}\sum_{\kappa=0}^{N_c-1} \mathcal{M}(\kappa+2,\kappa+1)\mathcal{M}(\kappa+1,\kappa) \\
&\quad - \left[\frac{1}{N_c}\sum_{\kappa=0}^{N_c-1} \mathcal{M}(\kappa+1,\kappa)\right]^2 \neq 0.
\end{aligned}
\label{eq:phase_induced_T2}
\end{equation}
Moreover, when static energetic disorder ($\sigma > 0$) and temporal phase averaging are present simultaneously, the total induced memory is \emph{non-additive}. We quantify this by the remainder
\begin{equation}
\Delta\mathcal{T}(d) = \bar{\mathcal{T}}_{\text{t-avg}}^{(\sigma)}(d) - \bar{\mathcal{T}}_{\text{t-avg}}^{(\sigma = 0)}(d) - \mathcal{T}^{(\sigma)}(d, 0), \qquad d \ge 2,
\label{eq:nonadditive}
\end{equation}
where $\mathcal{T}^{(\sigma)}(d,0)$ is the fixed-phase, disorder-averaged two-index tensor. To leading order, the disorder and the driving phase vary independently across the ensemble and their cross terms cancel, yet $\Delta\mathcal{T}(d) \neq 0$: the remainder is proportional to $\sigma^2$, like the disorder-only memory, and measures how differently the same disorder-induced deviation of a single step is carried forward by the phase-averaged dynamics and by the dynamics at a fixed phase (Appendix~\ref{app:scaling}). 

\emph{Frame dependence.} 
Under a time-dependent change of frame, $\rho \to \mathcal{R}^{-1}(t)\rho$ with $\mathcal{R}(t)$ a unitary superoperator, the two-index tensors of a single trajectory transform covariantly, $\mathcal{T}(n,k) \to \mathcal{R}^{-1}(t_n)\,\mathcal{T}(n,k)\,\mathcal{R}(t_k)$, which leaves their norms unchanged. The start-time average of Eq.~\eqref{eq:phase_avg_map}, however, combines maps whose frame factors $\mathcal{R}(t_\kappa)$ differ from one starting time to the next, so it does not commute with the change of frame. Phase-induced memory is therefore a property of the averaged dynamics \emph{in a specified frame}. Physically, the appropriate frame is set by the phase reference available for preparation and detection: a fixed (laboratory) reference when the driving phase varies across emitters or shots, as in the scenarios above, or a frame locked to the drive when every emitter is prepared and detected relative to its own driving phase.
For drives with several tones the averaged dynamics again depends on the frame, with two additional features that we leave to future work: averaging over the start time and averaging over the independent phases of the tones are no longer equivalent, and whether a change of frame can remove the phase-induced memory depends on the closed loops of couplings formed by the tones. 

\subsection{Generalized Outgoing Pseudoflux}
\label{subsec:theory_pseudoflux}
To diagnose localization and transport in driven lattice systems of $N$ sites, we extend the outgoing pseudoflux introduced in Ref~\cite{Anderson2026}.. In Liouville space, let $\mathcal{T}_{ii,jj}(n,k)$ denote the matrix element of $\mathcal{T}(n,k)$ transferring population $|j\rangle\langle j|$ at step $k$ to population $|i\rangle\langle i|$ at step $n$. For a two-index transfer tensor hierarchy, we define the \emph{two-time outgoing pseudoflux} from site $j$ at starting step $k$ and memory lag $d = n - k \ge 1$ as
\begin{equation}
f_j(d, k) = \sum_{i \neq j} \mathcal{T}_{ii,jj}(k + d, k),
\label{eq:two_time_pseudoflux}
\end{equation}
and the corresponding \emph{cumulative outgoing pseudoflux} up to memory lag $K$ as
\begin{equation}
F_j(K, k) = \sum_{d=1}^K f_j(d, k).
\label{eq:cum_two_time_pseudoflux}
\end{equation}
For stroboscopically sampled ($N_c = 1$) or temporally homogenized ($\bar{\mathcal{T}}_{\text{t-avg}}$) descriptions, these reduce to the one-index quantities $f_j(d) = \sum_{i\neq j} \bar{\mathcal{T}}_{ii,jj}(d)$ and $F_j(K) = \sum_{d=1}^K f_j(d)$.

As shown in Ref~\cite{Anderson2026}., trace preservation requires $\sum_i \bar{\mathcal{T}}_{ii,jj}(d) = \delta_{d,1}$, so the first step $f_j(1) = 1 - \bar{\mathcal{T}}_{jj,jj}(1) \ge 0$ measures the single-step population escaping site $j$, whereas the memory steps $f_j(d \ge 2) = -\bar{\mathcal{T}}_{jj,jj}(d)$ are predominantly negative and represent backflow returning population to its origin. When off-diagonal coherences dephase in the long-time limit---through environmental noise, static disorder averaging, or driving-phase homogenization---a localized steady-state population profile requires the cumulative outgoing pseudoflux to vanish asymptotically, $\lim_{K\to\infty} F_j(K) = 0$, whereas diffusive or delocalized transport yields $F_j(K) > 0$ over the diffusive timescale~\cite{Anderson2026}. Since all pseudoflux quantities for a particle prepared on site $j$ involve only the column $\mathcal{T}(n,k)\,|j\rangle\langle j|$ of the tensor hierarchy, we also use the norm of this column, $\|\mathcal{T}(d)\,|j\rangle\langle j|\|_F$, as a measure of the memory acting on the chosen initial state. For long chains, this column can be evaluated without constructing the $N^2 \times N^2$ superoperators (Appendix~\ref{app:column}).

\section{Physical Examples and Numerical Results}
\label{sec:examples}

\subsection{Periodically Driven Two-Level System: Phase-Induced Memory and the RWA}
\label{subsec:results_2ls}
We first benchmark our formalism on an ensemble of inhomogeneously broadened two-level emitters driven by a continuous-wave monochromatic field, a model directly relevant to optical-cycling molecular complexes in solution~\cite{Shin2024} and coherent optical spectroscopy~\cite{Noh2010}. For the $i$-th emitter with transition frequency $\omega_0^{(i)} = \bar{\omega}_0 + \delta\omega_i$ drawn from $\delta\omega_i \sim \mathcal{N}(0, \sigma^2)$, the lab-frame Hamiltonian is ($\hbar = 1$):
\begin{equation}
H^{(i)}(t) = \frac{\omega_0^{(i)}}{2} \sigma_z + V \cos(\omega_d t + \phi) \sigma_x,
\label{eq:h_2ls}
\end{equation}
where $\omega_d$ is the driving frequency, $V$ is the Rabi coupling strength, and $\phi$ is the initial driving phase. Each emitter undergoes Markovian relaxation governed by the Lindblad master equation $\dot{\rho}^{(i)} = -i[H^{(i)}(t), \rho^{(i)}] + \mathcal{L}_{\text{pd}}\rho^{(i)} + \mathcal{L}_{\text{sp}}\rho^{(i)}$, with pure dephasing dissipator $\mathcal{L}_{\text{pd}}\rho = 2\gamma_{\text{pd}}(\sigma_+ \sigma_- \rho \sigma_+ \sigma_- - \frac{1}{2}\{\sigma_+ \sigma_-, \rho\})$ and spontaneous emission dissipator $\mathcal{L}_{\text{sp}}\rho = 2\gamma_{\text{sp}}(\sigma_- \rho \sigma_+ - \frac{1}{2}\{\sigma_+ \sigma_-, \rho\})$.

Under the RWA, transforming to the frame rotating at $\omega_d$ and discarding the counter-rotating terms $\frac{V}{2}(e^{2i\omega_d t + i\phi}\sigma_+ + e^{-2i\omega_d t - i\phi}\sigma_-)$ yields a time-independent rotating-frame Bloch generator $\mathcal{L}_{\text{RWA}}^{(i)}$ that depends only on the detuning $\Delta_i = \omega_d - \omega_0^{(i)}$~\cite{Noh2010}. The rotating frame turns at the common driving frequency $\omega_d$, so the frame itself does not depend on the disorder; the inhomogeneity enters only through $\Delta_i$. Consequently, for a single emitter ($\sigma = 0$) in this frame, the RWA two-time map $\mathcal{M}^{\text{rot}}_{\text{RWA}}(n,k) = e^{\mathcal{L}_{\text{RWA}}(n-k)\Delta t}$ is independent of the starting step $k$. Averaging over starting times in Eq.~\eqref{eq:phase_avg_map} leaves this semigroup invariant, so in the rotating frame the RWA predicts identically zero phase-induced memory:
\begin{equation}
\bar{\mathcal{T}}_{\text{t-avg}}^{\text{RWA, rot}}(d) = 0 \quad \forall d \ge 2 \quad (\sigma = 0).
\label{eq:rwa_zero_memory}
\end{equation}
For the exact Hamiltonian~\eqref{eq:h_2ls}, the counter-rotating terms oscillate at $2\omega_d$ in the rotating frame; the single-step propagators $\mathcal{M}(\kappa+1,\kappa)$ vary across the cycle, and Eq.~\eqref{eq:phase_induced_T2} generates non-vanishing phase-induced memory even at $\sigma = 0$. In the rotating frame, phase-induced memory is thus a direct measure of the counter-rotating dynamics. 

In the lab frame the situation is different. With $\mathcal{R}(t)$ the rotation at $\omega_d$, the lab-frame RWA maps are $\mathcal{M}^{\text{lab}}_{\text{RWA}}(\kappa+d,\kappa) = \mathcal{R}(t_{\kappa+d})\,e^{\mathcal{L}_{\text{RWA}}d\Delta t}\,\mathcal{R}^{-1}(t_\kappa)$, i.e., maps from different starting times are rotated copies of one semigroup. Their average is a convex mixture of distinct Markovian evolutions and generates memory~\cite{Breuer2018} even within the RWA. Physically, the start-time average scrambles the phase of the Rabi coherence relative to the fixed reference, so the population$\to$coherence$\to$population pathways that drive Rabi oscillations must be restored by memory. Populations are identical in both frames; only the coherences, and therefore the transfer tensors, differ.

\begin{figure}[htbp]
\centering
\includegraphics[width=0.84\textwidth]{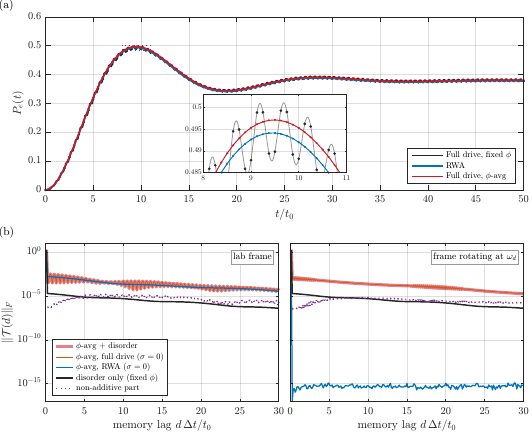}
\caption{Transfer tensor analysis of a periodically driven, inhomogeneously broadened two-level ensemble ($\bar\omega_0 = 10$, $\omega_d = 10.2$, $V = 0.5$, $\gamma_{\text{pd}} = \gamma_{\text{sp}} = 0.1$, $\sigma = 0.05$, $N_c = 8$). (a) Excited-state population $P_e(t)$ for the full semiclassical drive at fixed phase $\phi = 0$ (black), the RWA (blue), and the temporally homogenized full drive (red). Lines: TTM propagation [two-index block recursion, Eq.~\eqref{eq:tt_block_symmetry}, for the fixed-phase drive; one-index tensors for the RWA and the phase-averaged drive]; dots: direct propagation of the ensemble-averaged maps (every fourth point shown in the main panel). Inset: the region marked by the dotted box, showing the $2\omega_d$ counter-rotating micromotion of the fixed-phase trajectory about the smooth phase-averaged envelope; for the fixed-phase drive, the gray curve is direct propagation sampled at 64 points per driving period, and the black markers are the $N_c = 8$ TTM values, which lie on it. (b) Frobenius norm $\|\mathcal{T}(d)\|_F$ versus memory lag $d\,\Delta t / t_0$ (in driving periods), with the maps expressed in the lab frame (left) and in the frame rotating at $\omega_d$ (right): temporally homogenized ensemble (phase average + disorder, thick red), clean phase-averaged full drive (thin orange), clean phase-averaged RWA (blue), fixed-phase disorder-averaged two-index tensor $\mathcal{T}(d,0)$ (black; frame-independent norm), and the non-additive remainder $\|\Delta\mathcal{T}(d)\|_F$ of Eq.~\eqref{eq:nonadditive} (dotted purple). In the rotating frame the phase-averaged RWA sits at the double-precision floor for all $d \ge 2$; in the lab frame it carries memory comparable to that of the full drive.} 
\label{fig:2ls_driving}
\end{figure}

Figure~\ref{fig:2ls_driving} displays numerical results for $\bar{\omega}_0 = 10$, $\omega_d = 10.2$, $V = 0.5$, $\gamma_{\text{pd}} = \gamma_{\text{sp}} = 0.1$, and $\sigma = 0.05$, sampled with $N_c = 8$ sub-steps per driving period $t_0 = 2\pi/\omega_d$, which resolves the $2\omega_d$ counter-rotating component with four points per oscillation. The Gaussian average over detunings is evaluated with 20-point Gauss--Hermite quadrature, so that the comparisons below are free of sampling noise. As shown in Fig.~\ref{fig:2ls_driving}(a), propagating the ground-state initial condition $\rho(0) = |g\rangle\langle g|$ with the two-index block TTM recursion reproduces direct propagation of the ensemble-averaged maps to within $10^{-14}$ for the full drive, and the same holds for the RWA. The one-index temporally homogenized tensors $\bar{\mathcal{T}}_{\text{t-avg}}(d)$ (red) track the cycle-averaged population envelope while removing the $2\omega_d$ micromotion visible in the inset. The RWA slightly underestimates the excitation, consistent with the Bloch--Siegert shift $V^2/(4\bar\omega_0) \approx 0.006$, which moves the transition toward the blue-detuned drive. 

The memory content, in contrast, differs qualitatively [Fig.~\ref{fig:2ls_driving}(b)]. 
Static inhomogeneity alone produces a smooth, slowly decaying two-index memory, whose norm is the same in either frame. Temporal homogenization of the exact drive adds a phase-induced memory, which is present already for the clean emitter (thin orange curves). In the lab frame [Fig.~\ref{fig:2ls_driving}(b), left], this memory carries fast counter-rotating oscillations superimposed on slow Rabi-envelope beats, and for the width used here it exceeds the disorder-only memory. The phase-averaged RWA retains a lab-frame memory of similar size, but without the fast oscillations: in this frame, most of the phase-induced memory reflects the scrambling of the Rabi-coherence phase discussed above rather than counter-rotating dynamics. In the frame rotating at $\omega_d$ [Fig.~\ref{fig:2ls_driving}(b), right], the phase-averaged RWA memory lies at the double-precision floor for all $d \ge 2$, confirming Eq.~\eqref{eq:rwa_zero_memory}, while the full drive retains a smaller memory, the counter-rotating part of the phase-induced memory. Finally, phase averaging and disorder do not combine additively in either frame. The remainder $\|\Delta\mathcal{T}(d)\|_F$ of Eq.~\eqref{eq:nonadditive} is small compared with the total memory, but beyond a few driving periods it exceeds the disorder-only memory, so the influence of the static inhomogeneity on the phase-averaged dynamics cannot be inferred from the fixed-phase tensors alone. 

The relative size of the two contributions is not intrinsic; it reflects the width chosen here, $\sigma = 0.1\,V$. 
Both kinds of averaging generate memory in the same way (Appendix~\ref{app:scaling}). Each member of the ensemble evolves with single-step maps that deviate slightly from the ensemble mean. To leading order, the memory at lag $d$ is the correlation, across the ensemble, between these deviations at the first and the last step of the window. Averaging therefore acts on the emitter like a classical random field. For the inhomogeneous detunings this field is static; for the driving phase it varies periodically in time, following the drive waveform. The disorder-induced memory therefore scales as $\sigma^2$ and the phase-induced memory as $V^2$, and the two are comparable when $\sigma$ is of the order of $V$, with a prefactor that depends on the frame and the time step. For broader inhomogeneity the disorder-induced memory dominates. The non-additive remainder is also quadratic in $\sigma$ and vanishes as $V \to 0$; near the crossover it becomes comparable to the total memory, so that neither contribution can be treated as a perturbation of the other. 

Because phase averaging acts through the autocorrelation of the drive waveform, the phase-induced memory depends on the waveform as well as on its amplitude. Only the part of the drive that varies over the period contributes, each harmonic weighted by the response of the emitter to it over one time step, so that a static offset adds no memory and off-resonant or fast harmonics add little (Appendix~\ref{app:scaling}). For fine time steps, the memory per step scales as $\Delta t^2$. Which part of the drive counts as varying depends on the frame: the co-rotating component is static in the frame of the drive and generates no memory there, which is the RWA result above. 

\begin{figure}[htbp]
\centering
\includegraphics[width=0.98\textwidth]{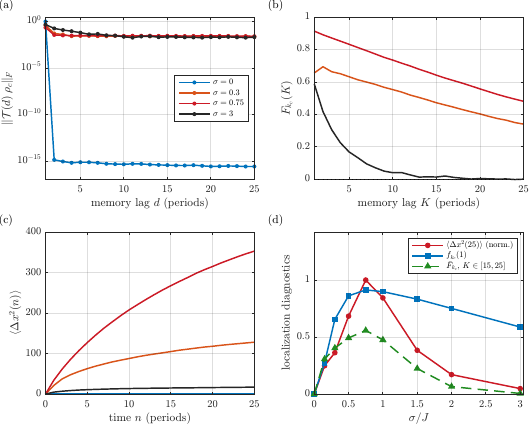}
\caption{Stroboscopic ($\Delta t = t_0$) transfer tensor and outgoing pseudoflux analysis of the Dunlap--Kenkre (DK) chain, Eq.~\eqref{eq:h_dk}, with $N = 161$ sites, $J = 1$, $\lambda = 1$, and $\eta = \lambda/\omega = 2.4048$ [$J_0(\eta) = 0$], for a particle prepared on the center site $k_c$. The clean chain is shown after averaging over $36$ initial phases $\phi_0$ (blue); the fixed-phase result is identical to numerical precision. For $\sigma > 0$, each of the $2000$ disorder realizations carries its own phase. (a) Memory acting on the initial state, $\|\mathcal{T}(k)\rho_c\|_F$ with $\rho_c = |k_c\rangle\langle k_c|$; for $\sigma = 0$, it drops from 1 at $k = 1$ to the numerical floor for $k \ge 2$; for $\sigma > 0$, its value beyond the first few lags is limited by the sampling error of the finite ensemble, which decreases as $R^{-1/2}$ (Appendix~\ref{app:numerics}). (b) Cumulative outgoing pseudoflux $F_{k_c}(k)$. (c) Mean squared displacement $\langle \Delta x^2(k) \rangle$. All data in (a)--(c) lie within the window in which the population within five sites of either chain end stays below $10^{-3}$. (d) Disorder scan: normalized spread $\langle \Delta x^2(25) \rangle$ (red circles), single-step outgoing pseudoflux $f_{k_c}(1)$ (blue squares), and cumulative pseudoflux averaged over $k \in [15, 25]$ (green triangles).}
\label{fig:dk_ttm}
\end{figure}

\subsection{Driven Lattice: Dunlap--Kenkre Localization and Disorder-Enhanced Spreading}
\label{subsec:results_dk}
As our second benchmark, we investigate coherent quantum transport on a 1D tight-binding chain of $N$ sites driven by a homogeneous AC electric field in the presence of static on-site disorder. The Hamiltonian is ($\hbar = 1$):
\begin{align}
H(t) &= J \sum_{j=1}^{N-1} \left(|j\rangle\langle j+1| + |j+1\rangle\langle j|\right) \nonumber \\
&\quad + \sum_{j=1}^N \left[\epsilon_j + (j - k_c)\,\lambda\cos(\omega t + \phi_0)\right] |j\rangle\langle j|,
\label{eq:h_dk}
\end{align}
where $J = 1$ is the nearest-neighbor hopping amplitude, $k_c = (N+1)/2$ is the center site, $\epsilon_j \sim \mathcal{N}(0, \sigma^2)$ is uncorrelated static Gaussian site-energy disorder, and $\lambda$ and $\omega$ are the AC driving amplitude and frequency. We tune the dimensionless driving ratio to the first root of the zeroth-order Bessel function, $\eta = \lambda/\omega \approx 2.4048$, with $\lambda = 1$, so that $\omega \approx 0.416$ and $t_0 = 2\pi/\omega \approx 15.1$. Note that this is a low-frequency drive, $\omega < J$; the bandwidth $4J$ exceeds the photon energy by an order of magnitude. Each realization is propagated exactly (up to a controlled time-step error) as described in Appendix~\ref{app:numerics}. Because a disordered wavepacket at this low drive frequency spreads by tens of sites within a few periods, we use a long chain ($N = 161$) and report only times for which the population within five sites of either end remains below $10^{-3}$, so that boundary reflections play no role.

In a clean lattice ($\sigma = 0$), Dunlap and Kenkre~\cite{Dunlap1986} showed that an AC field renormalizes the effective cycle-averaged hopping to $J_{\text{eff}} = J\,J_0(\lambda/\omega)$. For a nearest-neighbor chain this statement is exact at the stroboscopic level: in the Peierls gauge (Appendix~\ref{app:numerics}) each Bloch momentum $q$ accumulates the phase $2J\int_0^{t_0}\cos[q - A(t)]\,dt$ over one period, with $A(t) = \eta[\sin(\omega t + \phi_0) - \sin\phi_0]$, and this integral is proportional to $J_0(\eta)$ for every initial phase $\phi_0$. At the Bessel root, the one-period propagator is therefore the identity, $U(t_0, 0) = \mathbb{I}$, independent of $\phi_0$ (\emph{dynamical localization}); in a finite open chain this holds exactly away from the edges, and a particle prepared on $k_c$ returns to it after every period. When static disorder $\sigma > 0$ is introduced, random site-energy offsets break this exact cancellation. At moderate $\sigma$ the particle then spreads over many sites, whereas for $\sigma \gg J$ conventional Anderson localization~\cite{Anderson1958} confines it close to its initial site. In this single-particle, one-dimensional model the spreading remains bounded for every $\sigma > 0$ (see the discussion of long time results below), so the enhanced spreading at moderate disorder should not be confused with genuine delocalization.
The interplay of AC fields and disorder has been studied extensively. The field controls the localization length through the ratio of disorder strength to the Bessel-renormalized quasienergy bandwidth~\cite{Holthaus1995,Holthaus1996}, driving can enhance or suppress localization depending on its frequency~\cite{Martinez2006}, and correlated disorder adds further structure~\cite{Zhang2006}. Cold-atom experiments have recently mapped the localization phase diagram of a driven quasiperiodic chain and found clear departures from the Bessel-renormalized picture when the drive frequency is comparable to the bandwidth~\cite{Dotti2025}, the regime considered here; spatially nonuniform driving in photonic lattices~\cite{Schindler2025} and disorder in the drive amplitude itself~\cite{Tiwari2024} have also been explored. Our aim is therefore not to establish how disorder modifies dynamical localization as such, but to show how the memory structure of the transfer tensors diagnoses it.

Figure~\ref{fig:dk_ttm} presents our stroboscopic ($\Delta t = t_0$) transfer tensor and outgoing pseudoflux analysis of this crossover. For the clean chain, the exactness of $U(t_0,0) = \mathbb{I}$ has an immediate and instructive consequence: stroboscopically, every dynamical map acting on $\rho_c = |k_c\rangle\langle k_c|$ is the identity, both at fixed phase and after averaging over $\phi_0$. Hence $\mathcal{T}(1)\rho_c = \rho_c$, and $\mathcal{T}(k \ge 2)\rho_c$, $f_{k_c}(k)$,  $F_{k_c}(k)$, and $\langle\Delta x^2(k)\rangle$ are all numerically undetectable at every period [blue curves in Figs.~\ref{fig:dk_ttm}(a)--(c)]. Stroboscopic TTM thus describes clean DK localization as a trivially frozen map without memory: no population ever leaves the origin, so there is neither outgoing pseudoflux nor memory-mediated backflow. Phase averaging does not change this, because the stroboscopic map is phase independent. The intra-cycle excursion of the particle, which is large (Sec.~\ref{subsec:results_substrobo}), is invisible at this resolution. (In a finite open chain, the one-period propagator deviates from the identity only for sites within the reach of the intra-cycle excursion from the edges; the full-superoperator norm $\|\mathcal{T}(k)\|_F$, which sums over all initial sites, is dominated by such edge columns, which is why we monitor the memory acting on the bulk initial state instead.)

Static disorder changes the picture qualitatively. Already at $\sigma = 0.3$, the single-step outgoing pseudoflux jumps to $f_{k_c}(1) = 0.66$, and it reaches $0.91$ at $\sigma = 0.75$, reflecting the breakdown of single-period refocusing. The memory acting on the initial state becomes nonzero at all lags [Fig.~\ref{fig:dk_ttm}(a); beyond the first few lags, its plotted magnitude is set mainly by the sampling error of the finite ensemble, whereas the pseudoflux converges (Appendix~\ref{app:numerics})], and the cumulative pseudoflux remains large and positive throughout the observation window [$F_{k_c}(25) = 0.34$ and $0.48$ for $\sigma = 0.3$ and $0.75$; Fig.~\ref{fig:dk_ttm}(b)], accompanied by rapid wavepacket expansion to $\langle \Delta x^2(25) \rangle = 128$ and $353$ [Fig.~\ref{fig:dk_ttm}(c)]. At strong disorder ($\sigma = 3$, black), Anderson localization dominates: the single-step pseudoflux is $f_{k_c}(1) = 0.59$, but strong backflow at the next period ($f_{k_c}(2) = -0.17$) and beyond quenches the cumulative pseudoflux to $F_{k_c}(25) = 0.001$, and the spread grows only slowly, reaching $\langle \Delta x^2(25) \rangle = 17$.

Figure~\ref{fig:dk_ttm}(d) summarizes the non-monotonic competition across $\sigma/J \in [0, 3]$. All three metrics vanish in the clean limit, rise steeply with disorder, and decrease again in the Anderson regime. The spread after 25 periods peaks at $\sigma^* \approx 0.75$, the cumulative pseudoflux averaged over $k \in [15, 25]$ peaks at $0.75$, and the single-step pseudoflux $f_{k_c}(1)$ is maximal and nearly flat for $\sigma \approx 0.75$--$1$ and decays only slowly at larger $\sigma$ (to $0.59$ at $\sigma = 3$), because a strongly detuned neighbor still allows the particle to leave the origin within one period, and it is the subsequent backflow that localizes it. The cumulative pseudoflux, which incorporates this backflow, therefore tracks the spread far more faithfully than the single-step flux, in line with the role of $F_{k_c}$ as the localization diagnostic~\cite{Anderson2026}.

\emph{Long times.} 
Over the 25 periods of Fig.~\ref{fig:dk_ttm}, the spread at intermediate disorder is still growing, but it does not grow indefinitely. In this model a single particle moves in a band of finite width and is coupled to no bath, and the one-period propagator of each disorder realization has localized eigenstates, in line with rigorous~\cite{Ducatez2017} and numerical~\cite{Agarwal2017} results for driven Anderson models in one dimension. Genuine delocalization would require additional ingredients, such as noise or a bath, higher dimensions, several incommensurate drive frequencies~\cite{Agarwal2017}, or a drive that traces a topologically nontrivial loop in parameter space~\cite{Wauters2019}. Diagonalizing the one-period propagator of 401-site chains confirms that the infinite-time spread is finite for every disorder strength in Fig.~\ref{fig:dk_ttm}(d). It is largest near $\sigma \approx 0.75$, where it is about twice the value reached after 25 periods, and it shrinks at stronger disorder. The non-monotonic dependence on $\sigma$ in Fig.~\ref{fig:dk_ttm}(d) therefore reflects a non-monotonic Floquet localization length rather than a transition to transport. Correspondingly, for all disorder strengths we examined, the cumulative pseudoflux decays toward zero at long times, but its decay rate is itself non-monotonic in $\sigma$. The decay is slow at weak disorder, where the particle spreads slowly, slow again near $\sigma \approx 0.75$--$1$, where the localization length is largest, and fast at strong disorder. Resolving this decay requires a large ensemble: with $R$ realizations the long-time average of the maps has rank at most $RN$, so that for small $R$ the cumulative pseudoflux levels off at a value that depends on $R$.

\begin{figure}[htbp]
\centering
\includegraphics[width=0.98\textwidth]{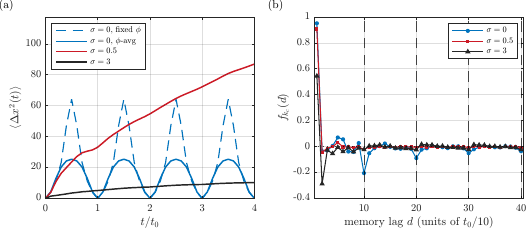}
\caption{Sub-stroboscopic ($N_c = 10$, $\Delta t = t_0/10$) transfer tensor analysis of the DK chain ($N = 101$, other parameters as in Fig.~\ref{fig:dk_ttm}). Temporal homogenization averages over the $N_c$ starting sub-steps (driving phases $2\pi\kappa/N_c$) and, for $\sigma > 0$, over $800$ disorder realizations. (a) Intra-cycle wavepacket spread $\langle \Delta x^2(t) \rangle$ over four driving periods for the clean chain at fixed phase $\phi_0 = 0$ (dashed blue) and after temporal homogenization (solid blue), and for $\sigma = 0.5$ (red) and $\sigma = 3$ (black). (b) Outgoing pseudoflux $f_{k_c}(d)$ of the temporally homogenized tensors versus memory lag $d$; vertical dashed lines mark period completions $d = 10, 20, 30, 40$.}
\label{fig:dk_substrobo}
\end{figure}

\subsection{Sub-Stroboscopic Dynamics: Intra-Cycle Breathing and Floquet Refocusing}
\label{subsec:results_substrobo}
Stroboscopic sampling captures net cycle-to-cycle transport but, as shown above, is blind to the dynamics within each driving period, and in particular to how the clean chain manages to return to its initial state. To resolve this, we discretize each period into $N_c = 10$ sub-steps ($\Delta t = t_0/10 \approx 1.51$) and construct the temporally homogenized tensors $\bar{\mathcal{T}}_{\text{t-avg}}(d)$ by averaging over the ten starting sub-steps. Over the four periods shown, the population within five sites of the ends of the $N = 101$ chain stays below $10^{-4}$.

Figure~\ref{fig:dk_substrobo}(a) shows the intra-cycle spread. For the clean chain at fixed phase $\phi_0 = 0$ (dashed blue), the wavepacket is far from stationary: it swells to $\langle \Delta x^2 \rangle = 64$ within each period before refocusing exactly onto $k_c$ at $t = m t_0$. After temporal homogenization (solid blue), the breathing amplitude is reduced to $25$. Each starting phase produces its own intra-cycle excursion, differing in size (peak spread between $23$ and $64$) and in timing. At every period completion, however, each starting phase refocuses exactly, so $\langle\Delta x^2(m t_0)\rangle = 0$ holds after averaging as well. With intermediate disorder ($\sigma = 0.5$, red), the initial expansion is comparable but the wavepacket fails to refocus and keeps spreading, reaching $\langle\Delta x^2(4t_0)\rangle = 87$. Strong disorder ($\sigma = 3$, black) clamps the spread from the outset ($\langle\Delta x^2(4t_0)\rangle = 10$).

The intra-cycle pseudoflux $f_{k_c}(d)$ [Fig.~\ref{fig:dk_substrobo}(b)] uncovers a distinction between the memory mechanisms of dynamical and Anderson localization that is invisible to stroboscopic observation:
\begin{enumerate}
\item In the clean dynamically localized chain ($\sigma = 0$, blue), the particle escapes rapidly at the first sub-step, $f_{k_c}(1) = 0.952$. During the remainder of the first period ($d = 2,\dots,9$), the memory pseudoflux is small and of both signs ($|f_{k_c}(d)| \le 0.07$). Precisely at the completion of each driving period ($d = 10, 20, 30, 40$), $f_{k_c}(d)$ exhibits sharp negative \emph{refocusing spikes}, $f_{k_c}(10) = -0.205$, $f_{k_c}(20) = -0.089$, $f_{k_c}(30) = -0.051$. These delayed, period-commensurate memory contributions drive the cumulative pseudoflux down, from $0.95$ after the first sub-step to $0.38$ after four periods, and are the TTM signature of Floquet refocusing.
\item For strong Anderson disorder ($\sigma = 3$, black), the destructive interference is caused by static site-energy offsets acting on the short timescale $1/\sigma \approx 0.33 \ll \Delta t$. Accordingly, after a smaller initial escape, $f_{k_c}(1) = 0.545$, the pseudoflux shows an \emph{immediate} backflow at the very next sub-step, $f_{k_c}(2) = -0.288$, followed by weaker negative contributions, so that $F_{k_c}$ falls to $0.04$ within the first period and to about zero after four ($-0.02$; $F_{k_c}$ is a signed sum and can slightly overshoot zero at finite lag); no period-commensurate structure is needed.
\item For intermediate disorder ($\sigma = 0.5$, red), the initial escape $f_{k_c}(1) = 0.908$ is as large as in the clean chain, and the immediate backflow is weak, $f_{k_c}(2) = -0.044$, because $\sigma$ is small compared with the bandwidth. At the same time, the phase accumulated from the disorder over one period, $\sigma t_0 \approx 7.6$, is large enough to scramble the relative phases required for refocusing: the period-completion spike is reduced from $f_{k_c}(10) = -0.205$ to $-0.022$. Population that escaped at $d = 1$ is therefore never coherently directed back to its origin, and the cumulative pseudoflux is still $F_{k_c}(40) = 0.75$ after four periods, while the wavepacket continues to spread [Fig.~\ref{fig:dk_substrobo}(a)]; at much longer times it saturates at the Floquet localization length (Sec.~\ref{subsec:results_dk}).
\end{enumerate}
Sub-stroboscopic transfer tensor analysis therefore shows that dynamical localization and Anderson localization achieve spatial confinement through different temporal memory structures: delayed, periodic refocusing at $d = mN_c$ versus immediate backflow at $d = 2$.

\emph{Choice of time step.} 
The refocusing spikes rely on a time step that divides the driving period. If $\Delta t/t_0 = a/q$ with $a > 1$, the sampled dynamics returns exactly only every $q$ steps, and for an incommensurate step it never returns exactly, only approximately. The clean chain remains localized in either case, because the one-period propagator is the identity for every phase, so the spread stays bounded. Numerically, the cumulative pseudoflux also stays between 0 and 1 and still vanishes asymptotically, but only over a time that grows rapidly with the number of distinct intra-cycle times that the steps visit. 
The reason follows in three steps. First, since $F_{k_c}(K) = 1 - \sum_{d=1}^{K}\bar{\mathcal{T}}_{k_ck_c,k_ck_c}(d)$ and $\sum_{d} \bar{\mathcal{T}}_{\text{t-avg}}(d)\, z^d = \mathbb{I} - \bar{\mathcal{M}}(z)^{-1}$ (Appendix~\ref{app:column}), the Abel limit gives $F_{k_c}(\infty) = \lim_{z\to 1^-}[\bar{\mathcal{M}}(z)^{-1}]_{k_ck_c,k_ck_c}$. Second, in the clean chain the averaged maps $\bar{\mathcal{M}}(n)$ do not decay but recur, so near $z = 1$ their generating function is dominated by the long-time average $\bar{\mathcal{M}}_\infty = \lim_{N\to\infty} N^{-1}\sum_{n<N}\bar{\mathcal{M}}(n)$, with $\bar{\mathcal{M}}(z) \simeq \bar{\mathcal{M}}_\infty/(1-z)$. If $\bar{\mathcal{M}}_\infty$ is invertible, $\bar{\mathcal{M}}(z)^{-1} \simeq (1-z)\,\bar{\mathcal{M}}_\infty^{-1} \to 0$ and $F_{k_c}(\infty) = 0$; the closer $\bar{\mathcal{M}}_\infty$ is to a singular map, the closer to $z=1$ this asymptotic form sets in, and hence the longer the pseudoflux takes to vanish. Third, $\bar{\mathcal{M}}_\infty$ is an average of the unitary maps at the intra-cycle times that the steps visit. An average over a few such maps remains well conditioned, whereas a dense sampling approaches the continuous average over the period, which strongly suppresses operators oscillating within a cycle and is therefore nearly singular. For the clean chain (41 sites, averaged over 40 driving phases), the smallest singular value of $\bar{\mathcal{M}}_\infty$ decreases from $2\times10^{-3}$ for $\Delta t = t_0/5$ to $10^{-6}$, $10^{-9}$, and $5\times10^{-11}$ for $t_0/10$, $t_0/20$, and $t_0/40$; an incommensurate step visits all intra-cycle times and gives a value comparable to the last. Correspondingly, the Abel-averaged $F_{k_c}$ after $10^4$ steps is $3\times10^{-3}$ for $\Delta t = t_0/10$, but $0.09$ and $0.07$ for $t_0/20$ and $2t_0/21$, and $0.16$ and $0.30$ for $t_0/40$ and the incommensurate step $\Delta t = (\sqrt{5}-1)\,t_0/12$; in the last two cases it falls below $10^{-3}$ only after about $10^8$ steps. A finite-time cumulative pseudoflux is therefore a reliable diagnostic of localization only for a coarse commensurate step, as used here, or for stroboscopic sampling; otherwise its convergence can be judged from the conditioning of the long-time average map, which, like the pseudoflux, does not depend on the initial state.

\section{Conclusion}
\label{sec:conclusion}
We have developed a two-time transfer tensor formalism $\mathcal{T}(n,k)$ for time-inhomogeneous and periodically driven open quantum systems, extending the principle that coarse-graining and resolution loss induce memory from spatial disorder~\cite{Anderson2026} to the temporal domain. For periodically driven systems sampled sub-stroboscopically ($\Delta t = t_0 / N_c$), the discrete block symmetry $\mathcal{T}(n+mN_c, k+mN_c) = \mathcal{T}(n,k)$ enables exact sub-cycle propagation to arbitrary times from only $N_c$ learned columns of dynamical maps. Averaging over the initial driving phase restores single-index TTI while generating phase-induced memory. For a driven two-level ensemble, this memory scales with the square of the Rabi coupling, whereas the disorder-induced memory scales with the square of the inhomogeneous width, so their relative size is set by the ratio of the inhomogeneous width to the Rabi coupling, with a crossover when the two are comparable, and the two combine non-additively. 
To leading order, both act on the emitter as a classical random field, static for the inhomogeneity and periodic for the driving phase, so that the memory magnitudes are set by $\sigma^2$ and by $V^2$ times the variance of the drive waveform over its period, respectively (Appendix~\ref{app:scaling}). 
Because start-time averaging does not commute with a time-dependent change of frame, the phase-induced memory must be specified together with a frame: in the frame rotating with the drive it isolates the counter-rotating dynamics and is erased by the rotating-wave approximation, whereas in the lab frame it largely survives that approximation. 
For the driven disordered tight-binding chain, stroboscopic transfer tensors describe the clean Dunlap--Kenkre chain as exactly frozen, while sub-stroboscopic tensors reveal that dynamical localization rests on delayed refocusing at integer multiples of the driving period, in contrast to the immediate backflow of Anderson localization. Intermediate disorder suppresses this refocusing, so that the particle spreads over a region much larger than in the clean or strongly disordered chain, producing a non-monotonic dependence of both the wavepacket spread and the cumulative outgoing pseudoflux on disorder strength. In this one-dimensional, single-particle model the spreading remains bounded: the Floquet states stay localized, and the cumulative pseudoflux decays toward zero at long times, at a rate that is itself non-monotonic in the disorder strength. Driven inhomogeneous ensembles are also an active experimental topic; in nitrogen-vacancy ensembles, for example, modulated drives that give all disorder realizations a common Floquet eigenframe suppress inhomogeneous dephasing~\cite{Satyajit2026}, offering a natural setting to probe phase-averaging-induced memory. 
A remaining limitation is that each two-time map is defined from a factorized system--bath state at the preparation time. When the drive has acted before the system is prepared, the system--bath correlations built up during the driving history enter as an additional, history-dependent inhomogeneous term, which can be addressed within the process-tensor formulation of TTM with interventions~\cite{PollockModi2018,Jorgensen2020,Gherardini2022}. 
Future work will extend this framework to multi-pulse nonlinear optical spectroscopy, drives with several tones, and initially correlated system--bath states~\cite{Buser2017}. 
 An interesting direction for future applications is to investigate the relationship between phase synchronization~\cite{settimo2026synchronization} and memory effects in periodically driven open quantum systems. In particular, the two-time transfer tensor framework could provide a means to examine how phase locking, coherence preservation, and environmental memory evolve under different driving conditions, and how these features are modified when averaging over different initial phases of the driving field.

\begin{acknowledgments}
This research was supported by the University of Nevada, Las Vegas. 
The authors used the AI assistants Claude (Opus 5.5, Anthropic) and Gemini (3.8 Flash, Google), the latter through agents in Google Antigravity, for literature searches, for developing and checking derivations (Appendices~\ref{app:column} and~\ref{app:scaling}), and for drafting and revising text. The authors directed this work, checked every reference against the original publication, verified the derivations, and reviewed and revised all text; they take full responsibility for the content of this article.
\end{acknowledgments}

\section*{Data Availability}
The data that support the findings of this article are openly available~\cite{FessahayeZenodo2026}, including the numerical data underlying Figs.~\ref{fig:2ls_driving}--\ref{fig:dk_substrobo} and the scaling analyses of Appendix~\ref{app:scaling}. The code used to generate the data and figures (Python and MATLAB) is available from the same reference. 

\appendix

\section{Tensor columns without full superoperators}
\label{app:column}
For a chain of $N$ sites, each superoperator has $N^4$ elements, so storing a hierarchy of $K$ tensors becomes prohibitive beyond $N \approx 40$; for $N = 161$ a single tensor occupies about 11~GB in double precision. All lattice quantities reported here, however, involve only the action of the tensors on the initial state $\rho_c$: the pseudoflux quantities $f_j(d)$ and $F_j(K)$ are sums of diagonal elements of the column $X_d \equiv \mathcal{T}(d)\rho_c$, and the memory measure is $\|X_d\|_F$. This appendix shows how these columns are obtained directly from the averaged maps. We consider a time-translationally invariant hierarchy (stroboscopic, or temporally homogenized), drop the subscript t-avg, and set $\bar{\mathcal{M}}(0) = \mathbb{I}$.

\emph{Why the order of the convolution matters.} The defining recursion, Eq.~\eqref{eq:tt_tavg}, places the tensors to the left of the maps. Applied to $\rho_c$, it reads
\begin{equation}
X_n = \bar{\mathcal{M}}(n)\rho_c - \sum_{m=1}^{n-1}\mathcal{T}(m)\,\big[\bar{\mathcal{M}}(n-m)\rho_c\big],
\label{eq:column_left}
\end{equation}
in which the earlier tensors act on the new matrices $\bar{\mathcal{M}}(n-m)\rho_c$ rather than on $\rho_c$. The earlier columns $X_m$ are then not sufficient, and the full superoperators are required. The same tensors, however, also satisfy the recursion with the opposite order. With the generating functions $\bar{\mathcal{M}}(z) = \sum_{n\ge0}\bar{\mathcal{M}}(n)z^n$ and $\mathcal{T}(z) = \sum_{n\ge1}\mathcal{T}(n)z^n$, Eq.~\eqref{eq:tt_tavg} states that $\bar{\mathcal{M}}(z) = \mathbb{I} + \mathcal{T}(z)\bar{\mathcal{M}}(z)$, that is, $\bar{\mathcal{M}}(z) = [\mathbb{I} - \mathcal{T}(z)]^{-1}$. A power series with leading term $\mathbb{I}$ has a two-sided inverse, so that also $\bar{\mathcal{M}}(z) = \mathbb{I} + \bar{\mathcal{M}}(z)\mathcal{T}(z)$, and
\begin{equation}
\bar{\mathcal{M}}(n) = \sum_{m=1}^{n}\mathcal{T}(m)\,\bar{\mathcal{M}}(n-m) = \sum_{m=1}^{n}\bar{\mathcal{M}}(n-m)\,\mathcal{T}(m),
\label{eq:two_orders}
\end{equation}
even though the maps and the tensors do not commute with one another. For example, both orders give $\mathcal{T}(3) = \bar{\mathcal{M}}(3) - \bar{\mathcal{M}}(1)\bar{\mathcal{M}}(2) - \bar{\mathcal{M}}(2)\bar{\mathcal{M}}(1) + \bar{\mathcal{M}}(1)^3$. Solving the right-hand form for $\mathcal{T}(n)$ and applying it to $\rho_c$ gives
\begin{equation}
X_n = \bar{\mathcal{M}}(n)\rho_c - \sum_{m=1}^{n-1} \bar{\mathcal{M}}(n-m)\,X_m , \qquad X_1 = \bar{\mathcal{M}}(1)\rho_c ,
\label{eq:column_recursion}
\end{equation}
in which the tensors enter only through the earlier columns. Every quantity is now an $N\times N$ matrix, and the averaged maps are needed only through their action on such matrices, $\bar{\mathcal{M}}(p)Y = \langle U_r(p)\,Y\,U_r(p)^\dagger\rangle_r$. The columns must be kept as full matrices even when only populations are of interest, because the averaged maps turn populations into coherences, which feed back into the populations of later columns.

\emph{Evaluating the maps.} For stroboscopic sampling, $U_r(p) = U_r^p$, where $U_r = W_r\,\mathrm{diag}(\mu_r)\,W_r^\dagger$ is the one-period propagator of realization $r$, with unimodular eigenvalues $\mu_{r,a}$. Then
\begin{equation}
U_r^{p}\,Y\,U_r^{\dagger p} = W_r\big[\Phi_r^{(p)}\circ(W_r^\dagger Y W_r)\big]W_r^\dagger, \qquad \big[\Phi_r^{(p)}\big]_{ab} = \big(\mu_{r,a}\,\mu_{r,b}^{*}\big)^{p},
\label{eq:floquet_phase}
\end{equation}
where $\circ$ denotes the elementwise product: in the eigenbasis of $U_r$, propagation over $p$ periods multiplies each matrix element by a phase. Because Eq.~\eqref{eq:column_recursion} is linear, its sum can be accumulated in each eigenbasis. Each column $X_m$ is transformed once into every eigenbasis when it is produced, $B_{r,m} = W_r^\dagger X_m W_r$. For each new $n$, the phase-weighted sum $\sum_m \Phi_r^{(n-m)}\circ B_{r,m}$ is formed elementwise, and a single back-transformation per realization completes the step; the term $\bar{\mathcal{M}}(n)\rho_c$ enters the same sum as $m = 0$, with $B_{r,0} = W_r^\dagger\rho_c W_r$ and the opposite sign. We obtain $W_r$ from the complex Schur decomposition, which returns a unitary eigenbasis even for degenerate eigenvalues, as for the clean chain at the $J_0$ root, where $U_r = \mathbb{I}$.

For temporally homogenized sub-stroboscopic maps, the average runs over realizations and starting sub-steps $\kappa$. A lag of $qN_c + s$ steps consists of $q$ whole periods followed by $s$ sub-steps, $U_{r,\kappa}(qN_c + s) = U_{r,\kappa}(s)\,F_{r,\kappa}^{\,q}$, where $F_{r,\kappa}$ is the one-period propagator starting at sub-step $\kappa$. The whole periods become phases in the eigenbasis of $F_{r,\kappa}$, as in Eq.~\eqref{eq:floquet_phase}, and the partial propagator $U_{r,\kappa}(s)$ is applied in the back-transformation, which is shared by all terms with the same $s = (n-m) \bmod N_c$.

\emph{Cost and checks.} For $R$ realizations and $K$ tensors, the elementwise accumulation costs $RK^2N^2$ operations and the basis changes cost $RKN^3$, compared with $K^2N^6$ for the full recursion with $N^2\times N^2$ superoperators; the transformed columns occupy $RKN^2$ elements instead of $KN^4$. We verified that Eq.~\eqref{eq:column_recursion} reproduces the columns of the full recursion to $10^{-14}$ on small chains, and that, for $N = 41$, the sub-stroboscopic pseudoflux and mean squared displacement agree with an independent full-superoperator implementation to within $10^{-4}$ and $10^{-3}$, respectively.

\section{Numerical details}
\label{app:numerics}
\emph{Use of AI tools.} The Python and MATLAB code that computes the dynamical maps, transfer tensors, pseudoflux and figures was written and debugged with the assistance of AI coding agents, Claude (Opus 5.5, Anthropic) and Gemini (3.8 Flash, Google, in Google Antigravity). The authors specified the models, parameters and algorithms, and verified the code by convergence tests in the time step and number of realizations, by comparison of transfer-tensor propagation with direct propagation of the averaged maps, and by comparison with independent implementations, including the full superoperator recursion (Appendix~\ref{app:column}). 

\emph{Two-level ensemble.} Single-step maps $\mathcal{M}^{(i)}(\kappa+1,\kappa)$ are obtained by a midpoint exponential integrator of the time-dependent Lindblad generator with 25 micro-steps per $\Delta t$. The RWA maps are exact exponentials of the rotating-frame generator; lab-frame maps follow from the exact transformation $\mathcal{R}(t)X = U X U^\dagger$ with $U = \mathrm{diag}(1, e^{-i\omega_d t})$ in the basis $(|g\rangle, |e\rangle)$, applied both to the RWA maps and, in the opposite direction, to the full-drive maps for the right panel of Fig.~\ref{fig:2ls_driving}(b). 

\emph{Driven chain.} Propagators are computed in the vector-potential (velocity) gauge, obtained from Eq.~\eqref{eq:h_dk} by the site-diagonal transformation $G(t) = \mathrm{diag}(e^{-i(j-k_c)A(t)})$ with $A(t) = \int_0^t \lambda\cos(\omega t' + \phi_0)\,dt'$. This removes the potential that grows linearly with position and moves the field into Peierls phases on the hopping, $H'(t) = J\sum_j (e^{-iA(t)}|j\rangle\langle j+1| + \text{h.c.}) + \sum_j \epsilon_j |j\rangle\langle j|$; we refer to it as the Peierls gauge. The Hamiltonian then remains bounded on long chains. We use the midpoint rule with $400$ steps per driving period; each step is evaluated exactly as $D(A_m)\,e^{-iH_s\delta t}\,D(A_m)^\dagger$ with $D(A) = \mathrm{diag}(e^{i(j-k_c)A})$ and $H_s$ the static Hamiltonian. The length-gauge propagators of Eq.~\eqref{eq:h_dk} follow exactly as $U(t_2,t_1) = G(t_2)U'(t_2,t_1)G(t_1)^\dagger$, where $G(t)$ reduces to the identity at multiples of $t_0$. Halving the time step changes the one-period propagator by at most $2\times10^{-4}$ in any element and the column acting on $|k_c\rangle$ by less than $10^{-4}$. For the stroboscopic disorder scan, we use $R = 2000$ realizations per value of $\sigma$, realization $r$ carrying the initial phase $\phi_0 = 2\pi r/R$. Averages over a finite ensemble carry a sampling error that decreases as $R^{-1/2}$. For $\sigma > 0$ this error sets the column norm $\|\mathcal{T}(d)\rho_c\|_F$ beyond the first few lags: at $d = 25$ it decreases from $0.08$ to $0.04$ and $0.02$ for $R = 120$, $480$ and $1920$ ($\sigma = 0.3$ and $3$). The pseudoflux converges faster: at $R = 1920$, $F_{k_c}(25)$ differs between independent sets of realizations by less than $0.005$, and its mean changes by less than $0.005$ between $R = 480$ and $1920$; for the sub-stroboscopic study, $R = 800$ realizations are combined with the $N_c = 10$ starting sub-steps. Boundary effects are monitored through the population within five sites of either chain end, which stays below $10^{-3}$ ($N = 161$, $k \le 25$) and $10^{-4}$ ($N = 101$, $t \le 4t_0$) for all data shown.

\section{Scaling of averaging-induced memory}
\label{app:scaling}
This appendix shows how the memory generated by averaging scales with the inhomogeneous width $\sigma$, the drive amplitude $V$ and waveform, and the time step, and explains the crossover and the non-additivity reported in Sec.~\ref{subsec:results_2ls}. Numbers refer to the two-level model of Sec.~\ref{subsec:results_2ls} in the lab frame unless stated otherwise.

\subsection{Leading-order covariance formula}
Consider a family of realizations $a$, each evolving under a Markovian time-dependent generator, with single-step maps $\mathcal{S}_a(j) \equiv \mathcal{M}_a(j+1,j)$ so that $\mathcal{M}_a(n,k) = \mathcal{S}_a(n-1)\cdots\mathcal{S}_a(k)$. The label $a$ is the disorder realization $\xi$ (energetic or spatial averaging), the starting step $\kappa$ (temporal homogenization, for which $\mathcal{S}_\kappa(j) = \mathcal{S}(\kappa+j)$), or both. Writing $\mathcal{S}_a(j) = \bar{\mathcal{S}}(j) + \delta\mathcal{S}_a(j)$ with $\langle\delta\mathcal{S}_a(j)\rangle_a = 0$, the averaged map $\bar{\mathcal{M}}(n,k) = \langle\mathcal{M}_a(n,k)\rangle_a$ is, to second order in $\delta\mathcal{S}$, the product of mean steps plus one term for every pair of steps $i<j$, in which $\delta\mathcal{S}_a(j)$ and $\delta\mathcal{S}_a(i)$ replace the mean steps and are averaged together. Grouping these pairs by their later member and comparing with Eq.~\eqref{eq:tt_two_index_recursion} gives $\mathcal{T}(n,n-1) = \bar{\mathcal{S}}(n-1)$ and
\begin{equation}
\mathcal{T}(n,k) = \big\langle \delta\mathcal{S}_a(n-1)\,\bar{\mathcal{S}}(n-2)\cdots\bar{\mathcal{S}}(k+1)\,\delta\mathcal{S}_a(k)\big\rangle_a + O(\delta\mathcal{S}^3), \qquad n-k \ge 2 .
\label{eq:app_cov}
\end{equation}
The memory at lag $d = n-k$ is therefore the ensemble covariance between the deviations of the first and last steps of the window from their mean, propagated through the mean dynamics in between. For $d = 2$, Eq.~\eqref{eq:app_cov} is exact and reduces to Eq.~\eqref{eq:two_time_cov_memory}. For temporal homogenization, the mean step $\bar{\mathcal{S}}$ is the same at every step, and
\begin{equation}
\bar{\mathcal{T}}_{\text{t-avg}}(d) = \big\langle \delta\mathcal{S}_{\kappa,\xi}(d-1)\,\bar{\mathcal{S}}^{\,d-2}\,\delta\mathcal{S}_{\kappa,\xi}(0)\big\rangle_{\kappa,\xi} + O(\delta\mathcal{S}^3).
\label{eq:app_cov_tavg}
\end{equation}
For $d \le 16$, Eqs.~\eqref{eq:app_cov} and \eqref{eq:app_cov_tavg} reproduce the exact tensors to within a relative error of $3.4\times10^{-3}$ for disorder alone ($\sigma = 0.05$, fixed phase) and $7.6\times10^{-2}$ for the clean phase average ($V = 0.5$). The corrections grow with the deviation accumulated over the window; for disorder they are of relative order $(\sigma t_d)^2$, reaching $43\%$ at $\sigma t_d = 0.6$.

\subsection{Fine time steps: averaging as a classical random field}
Let the Hamiltonian of realization $a$ differ from the mean by $\delta H_a(t) = \eta_a(t)\,X$, so that its generator is $\bar{\mathcal{L}}(t) - i\,\mathrm{ad}_{\delta H_a(t)}$ with $\mathrm{ad}_X\rho = [X,\rho]$ and $\mathrm{ad}_X^2\rho = [X,[X,\rho]]$. When $\Delta t$ is short compared with all system time scales, $\delta\mathcal{S}_a(j) \approx -i\Delta t\,\eta_a(t_j)\,\mathrm{ad}_X$, and Eq.~\eqref{eq:app_cov_tavg} becomes (Eq.~\eqref{eq:app_cov} similarly, with the fixed-phase mean propagation in between)
\begin{equation}
\bar{\mathcal{T}}_{\text{t-avg}}(d) \approx -\Delta t^2\, C(t_{d-1})\;\mathrm{ad}_X\,\bar{\mathcal{S}}^{\,d-2}\,\mathrm{ad}_X, \qquad C(\tau) = \langle \eta_a(t+\tau)\,\eta_a(t)\rangle_a ,
\label{eq:app_noise}
\end{equation}
which has the form of the second-order memory kernel generated by a classical random field whose values at times separated by $\tau$ have correlation $C(\tau)$, sampled at intervals $\Delta t$. Both kinds of averaging act in this way and differ only in $C(\tau)$:
\begin{itemize}
\item \emph{Energetic disorder}: realization $\xi$ has transition energy $\bar E_0 + \sigma\xi$, where $\xi$ is a Gaussian variable with zero mean and unit variance. Then $\eta = \sigma\xi$ and $X = \sigma_z/2$, so $C(\tau) = \sigma^2$ is constant: the field is static. Independent site energies in a lattice, $\epsilon_j = \sigma\xi_j$, give a sum of such terms, $\sigma^2\sum_j \mathrm{ad}_{n_j}\cdots\mathrm{ad}_{n_j}$.
\item \emph{Driving-phase averaging}: for a drive $V f(\omega_d t + \phi)\,\sigma_x$ with $f(x) = \sum_n c_n e^{inx}$, $\eta = V[f - \bar f]$ and $X = \sigma_x$, so $C(\tau) = V^2\sum_{n\neq0}|c_n|^2 e^{in\omega_d\tau}$ is the autocorrelation of the waveform. It is periodic and averages to zero over a cycle, and $C(0) = V^2\,\mathrm{Var}(f)$, where $\mathrm{Var}(f) = \overline{f^2} - \bar f^{\,2} = \sum_{n\neq0}|c_n|^2$ is the variance of the waveform over one period.
\end{itemize}
Four consequences follow.

\emph{Magnitude.} To leading order in $\sigma$ and $V$, the two-step memory obeys
\begin{equation}
\lim_{\Delta t\to 0}\frac{\|\mathcal{T}_{\text{dis}}(2)\|_F}{\Delta t^2} = \sqrt{2}\,\sigma^2, \qquad \lim_{\Delta t\to 0}\frac{\|\bar{\mathcal{T}}_{\text{ph}}(2)\|_F}{\Delta t^2} = \sqrt{32}\,V^2\,\mathrm{Var}(f),
\label{eq:app_magnitude}
\end{equation}
where the limit is taken at fixed driving period ($N_c \to \infty$). The prefactors are $\sqrt{2} = \|\mathrm{ad}_{\sigma_z/2}^2\|_F$ and $\sqrt{32} = \|\mathrm{ad}_{\sigma_x}^2\|_F$: for a qubit, $\mathrm{ad}_X$ has eigenvalues $0$, $0$ and $\pm\Delta_X$, with $\Delta_X$ the difference between the two eigenvalues of $X$, so that $\|\mathrm{ad}_X^2\|_F = \sqrt{2}\,\Delta_X^2$, with $\Delta_X = 1$ for $\sigma_z/2$ and $2$ for $\sigma_x$. For $f = \cos$, for which $\mathrm{Var}(f) = 1/2$, and $N_c = 128$, the computed $\|\mathcal{T}(2)\|_F/\Delta t^2$ is $1.412\,\sigma^2$ and $2.817\,V^2$, within $0.2\%$ and $0.4\%$ of these limits. Both memories scale as $\Delta t^2$, i.e.\ as $N_c^{-2}$ at fixed period, so $\mathcal{T}(d)/\Delta t^2$ converges to a continuous memory kernel.

\emph{Crossover.} The two contributions are equal when $\sigma^* = V\,[\mathrm{Var}(f)\,\|\mathrm{ad}_{\sigma_x}^2\|_F/\|\mathrm{ad}_{\sigma_z/2}^2\|_F]^{1/2}$, which is $\sqrt{2}\,V$ for $f = \cos$. In the frame rotating at $\omega_d$, only the counter-rotating field $\tfrac{V}{2}[\cos(2\omega_d t)\sigma_x - \sin(2\omega_d t)\sigma_y]$ varies across starting phases, which gives $\sigma^* = 0.93\,V$. The computed crossover approaches these limits as the step is refined: $\sigma^*/V = 1.08$, $1.32$, $1.39$, $1.41$ and $1.41$ for $N_c = 8$, $16$, $32$, $64$ and $128$ in the lab frame, and $0.63$, $0.90$ and $0.93$ for $N_c = 8$, $32$ and $128$ in the rotating frame. At $N_c = 8$, the disorder memory is within $3\%$ of its fine-step value, whereas the phase memory is reduced to $56\%$ (lab frame) and $46\%$ (rotating frame), because the emitter precesses by $\bar\omega_0\Delta t \approx 0.8$ rad, and the counter-rotating field turns by $2\omega_d\Delta t = \pi/2$, within a single step.

\emph{Lag dependence.} For disorder $C$ is constant, so the memory decays only through the mean propagation $\bar{\mathcal{S}}^{\,d-2}$, i.e.\ on the relaxation time of the averaged emitter and, to leading order, independently of $\sigma$: $\|\mathcal{T}(d,0)\|_F$ decays at a rate of $\approx 0.22$ per unit time, close to the Bloch relaxation rate $\gamma_{\text{sp}} + \gamma_{\text{pd}} = 0.2$. At $N_c = 8$, where the lag $d = 256$ spans 32 driving periods, $\|\mathcal{T}(256,0)\|_F/\|\mathcal{T}(2,0)\|_F = 0.0145$ and $0.0134$ for $\sigma = 0.025$ and $0.1$. For phase averaging, $C$ oscillates, and $\|\bar{\mathcal{T}}_{\text{t-avg}}(d)\|_F$ oscillates in $d$ (here with a period of four steps, half a driving period) under the same relaxation envelope. Comparisons based on norms averaged over many lags therefore give a different crossover from the $d = 2$ value: the ratio of the medians over 50 periods of disorder-only to phase-induced memory is $0.66$ at $\sigma = 0.54$ and $1.72$ at $\sigma = 1$ ($V = 0.5$).

\emph{Waveform and finite steps.} Expanding $\delta\mathcal{S}_\kappa$ to first order in $V$, but exactly in $\Delta t$, gives $\delta\mathcal{S}_\kappa = V\sum_n c_n e^{in\omega_d\kappa\Delta t}\,Y_n$ with
\begin{equation}
Y_n = \int_0^{\Delta t} e^{\mathcal{L}_0(\Delta t-s)}\,e^{in\omega_d s}\,(-i\,\mathrm{ad}_{\sigma_x})\,e^{\mathcal{L}_0 s}\,ds,
\end{equation}
where $\mathcal{L}_0$ is the undriven generator. Averaging over $\kappa$ in Eq.~\eqref{eq:app_cov_tavg} then yields
\begin{equation}
\bar{\mathcal{T}}_{\text{t-avg}}(d) = V^2 \sum_{n}\ \sum_{m}\ c_n c_m\, e^{in\omega_d(d-1)\Delta t}\; Y_n\, e^{\mathcal{L}_0 (d-2)\Delta t}\, Y_m + O(V^3),
\label{eq:app_harmonic}
\end{equation}
where $n$ runs over all integers that are not multiples of $N_c$ and $m$ over all integers for which $n + m$ is a multiple of $N_c$. Harmonics with $n$ a multiple of $N_c$ are the same at every starting step and enter only the mean step $\bar{\mathcal{S}}$, and the average over $\kappa$ of $e^{2\pi i (n+m)\kappa/N_c}$ vanishes unless $n + m$ is a multiple of $N_c$. Only the varying part of the waveform contributes. Each harmonic is weighted by $Y_n$, the response of the emitter at frequency $n\omega_d$ filtered over one step, and harmonics $n$ and $n + N_c$ are indistinguishable on the time grid (aliasing). Equation~\eqref{eq:app_harmonic} reduces to Eq.~\eqref{eq:app_noise} when $n\omega_d\Delta t$ and $\bar\omega_0\Delta t$ are small. For the waveforms discussed in Sec.~\ref{subsec:results_2ls} ($V = 0.5$, $N_c = 8$), it reproduces the exact $\|\mathcal{T}(2)\|_F$ to within $1\%$: $2.349\times10^{-3}$ (exact $2.345\times10^{-3}$) for $\cos$ with or without a static offset, $4.687\times10^{-4}$ ($4.681\times10^{-4}$) for $\cos 2x$, $3.383\times10^{-3}$ ($3.360\times10^{-3}$) for a square wave, and $4.246\times10^{-4}$ ($4.207\times10^{-4}$) and $1.421\times10^{-4}$ ($1.421\times10^{-4}$) for pulse trains with duty cycles $1/4$ and $1/8$. Which part of the drive varies across starting phases depends on the frame. In the rotating frame the co-rotating field is static and drops out, which gives Eq.~\eqref{eq:rwa_zero_memory}. In the lab frame, the RWA field still varies across starting phases, which accounts for the lab-frame RWA memory in Fig.~\ref{fig:2ls_driving}(b), left.

\subsection{Combined averaging and non-additivity}
With disorder and starting phase averaged together, $a = (\kappa,\xi)$ and $\delta\mathcal{S}_{\kappa,\xi} = \delta\mathcal{S}^{\text{ph}}_\kappa + \sigma\xi\,\partial\mathcal{S}_\kappa + O(\sigma^2)$, where $\partial\mathcal{S}_\kappa$ is the derivative of the step map with respect to the transition frequency. Because $\langle\xi\rangle = 0$, the mixed second-order terms vanish, and the averaged tensor separates as
$\bar{\mathcal{T}}^{(\sigma)}_{\text{t-avg}}(d) = \bar{\mathcal{T}}^{(\sigma=0)}_{\text{t-avg}}(d) + \sigma^2 D_{\text{t-avg}}(d) + O(\sigma^4)$.
Here $D_{\text{t-avg}}$ collects all terms quadratic in $\sigma$, to all orders in $V$: the covariance of $\partial\mathcal{S}_\kappa$ averaged over $\kappa$ and propagated by the phase-averaged dynamics. The fixed-phase tensor is instead $\mathcal{T}^{(\sigma)}(d,0) = \sigma^2 D_{\text{fixed}}(d) + O(\sigma^4)$, the same deviation propagated by the phase-locked driven dynamics. The non-additive remainder of Eq.~\eqref{eq:nonadditive} is therefore
\begin{equation}
\Delta\mathcal{T}(d) = \sigma^2\,[D_{\text{t-avg}}(d) - D_{\text{fixed}}(d)] + O(\sigma^4).
\label{eq:app_nonadd}
\end{equation}
It is quadratic in $\sigma$, like the disorder memory itself, and vanishes as $V \to 0$, when phase-locked and phase-averaged dynamics coincide. Numerically, $\|\Delta\mathcal{T}(2)\|_F/\sigma^2 = 2.14\times10^{-4}$ to within $0.6\%$ for $\sigma$ between $0.0125$ and $1$. The ratio $\|\Delta\mathcal{T}(d)\|_F/\|\mathcal{T}(d,0)\|_F$ is independent of $\sigma$ at each lag: about $0.05$ for $d \le 16$, it exceeds unity beyond about six driving periods, and its median over 50 periods is $3.24$, $3.28$ and $3.30$ for $\sigma = 0.025$, $0.05$ and $0.1$. At short lags it grows with the drive, $0.004$, $0.009$, $0.019$, $0.050$ and $0.185$ for $V = 0.0625$, $0.125$, $0.25$, $0.5$ and $1$ ($\sigma = 0.02$, median over $d \le 16$). The dependence on the drive changes with lag. Within the first half period ($d \le 4$), $\|\Delta\mathcal{T}(d)\|_F$ is proportional to $\sigma^2 V$: the fixed-phase tensor contains a term linear in $V$, which arises because the drive at that particular phase modifies how each step responds to the detuning, and this term averages to zero over starting phases. Over two to four periods ($d = 16$--$32$), $\|\Delta\mathcal{T}(d)\|_F$ grows approximately as $\sigma^2 V^2$, as the difference between propagation by the phase-locked and by the phase-averaged dynamics accumulates. At longer lags, where $V$ also sets the phase of the Rabi oscillations, no single power of $V$ applies. The non-additivity therefore introduces no new scaling with $\sigma$. It reflects that a disorder-induced deviation is propagated by different mean dynamics in the two cases, a difference that accumulates with lag. Near the crossover it is comparable to the total memory: at $\sigma = 0.54$ the median of $\|\Delta\mathcal{T}(d)\|_F/\|\bar{\mathcal{T}}^{(\sigma)}_{\text{t-avg}}(d)\|_F$ is $1.3$.

\subsection{Scope}
Table~\ref{tab:scaling} summarizes these results. They hold while the deviation accumulated over the lag window is small, which for the two-level ensemble is the case for $\sigma, V \lesssim 1$ at the short lags that dominate the memory. They do not apply to the stroboscopic Dunlap--Kenkre chain of Sec.~\ref{subsec:results_dk}: there $\Delta t = t_0 \approx 15/J$, so $\sigma\Delta t \gg 1$ for every disorder strength in Fig.~\ref{fig:dk_ttm}, the memory is not quadratic in $\sigma$, and its non-monotonic dependence on $\sigma$ is a non-perturbative effect.

\begin{table}[htbp]
\centering
\caption{Leading-order scaling of averaging-induced memory for the driven two-level ensemble. $\Delta t$ is the time step; $\|\cdot\|$ is the Frobenius norm.}
\label{tab:scaling}
\small
\setbox0\vbox{\tabskip0pt\halign{\strut#\hfil\tabskip1.4em&#\hfil&#\hfil&#\hfil\tabskip0pt\cr
\noalign{\hrule\vskip1.5pt\hrule\vskip2pt}
Averaging & Correlation $C(\tau)$ & $\|\mathcal{T}(2)\|$ at fine steps & Lag dependence\cr
\noalign{\vskip2pt\hrule\vskip2pt}
Energetic disorder & $\sigma^2$ (static) & $\sqrt{2}\,\sigma^2\Delta t^2$ & relaxation of the mean dynamics\cr
Driving phase & $V^2\sum_{n\neq0}|c_n|^2e^{in\omega_d\tau}$ & $\sqrt{32}\,V^2\,\mathrm{Var}(f)\,\Delta t^2$ & oscillates at drive harmonics\cr
Both together & sum of the above & sum $+\;\|\Delta\mathcal{T}\| \propto \sigma^2$ & $\|\Delta\mathcal{T}\|/\|\mathcal{T}_{\text{dis}}\|$ grows with lag\cr
\noalign{\vskip2pt\hrule\vskip1.5pt\hrule}
}}
\centerline{\box0}
\end{table}

\bibliography{references}

@book{Breuer2002,
  title = {The Theory of Open Quantum Systems},
  author = {Breuer, Heinz-Peter and Petruccione, Francesco},
  year = {2002},
  publisher = {Oxford University Press},
  address = {Oxford}
}

@article{Cerrillo2014,
  title = {Non-{M}arkovian Dynamical Maps: Numerical Processing of Open Quantum Trajectories},
  author = {Cerrillo, Javier and Cao, Jianshu},
  journal = {Phys. Rev. Lett.},
  volume = {112},
  issue = {11},
  pages = {110401},
  year = {2014},
  doi = {10.1103/PhysRevLett.112.110401}
}

@article{Anderson2026,
  title = {Transfer tensor analysis of localization in the {A}nderson and {A}ubry-{A}ndr{\'e}-{H}arper models},
  author = {Anderson, Michelle C. and Chuang, Chern},
  journal = {Phys. Rev. Res.},
  volume = {8},
  issue = {1},
  pages = {013248},
  year = {2026},
  doi = {10.1103/y49y-7vhp}
}

@article{Dunlap1986,
  title = {Dynamic localization of a charged particle moving under the influence of an electric field},
  author = {Dunlap, D. H. and Kenkre, V. M.},
  journal = {Phys. Rev. B},
  volume = {34},
  issue = {6},
  pages = {3625--3633},
  year = {1986},
  doi = {10.1103/PhysRevB.34.3625}
}

@article{Anderson1958,
  title = {Absence of Diffusion in Certain Random Lattices},
  author = {Anderson, P. W.},
  journal = {Phys. Rev.},
  volume = {109},
  issue = {5},
  pages = {1492--1505},
  year = {1958},
  doi = {10.1103/PhysRev.109.1492}
}

@article{Follia2026,
  title = {Control of memory effects in a spin-boson system by periodic driving},
  author = {Follia, Pietro Marco and Vacchini, Bassano and Breuer, Heinz-Peter},
  journal = {Phys. Rev. A},
  volume = {113},
  issue = {2},
  pages = {022203},
  year = {2026},
  doi = {10.1103/PhysRevA.113.022203}
}

@article{Kananenka2016,
  title = {Accurate Long-Time Mixed Quantum-Classical {L}iouville Dynamics via the Transfer Tensor Method},
  author = {Kananenka, Alexei A. and Hsieh, Chang-Yu and Cao, Jianshu and Geva, Eitan},
  journal = {J. Phys. Chem. Lett.},
  volume = {7},
  issue = {23},
  pages = {4809--4814},
  year = {2016},
  doi = {10.1021/acs.jpclett.6b02390}
}

@article{Buser2017,
  title = {Initial system-environment correlations via the transfer-tensor method},
  author = {Buser, Maximilian and Cerrillo, Javier and Schaller, Gernot and Cao, Jianshu},
  journal = {Phys. Rev. A},
  volume = {96},
  issue = {6},
  pages = {062122},
  year = {2017},
  doi = {10.1103/PhysRevA.96.062122}
}

@article{Gao2026,
  title = {Extracting Transfer Tensors from Stochastic {S}chr{\"o}dinger Equation Trajectories: A Novel Approach to Dissipative Quantum Dynamics},
  author = {Gao, Siyuan and Lan, Zhihao and Liang, WanZhen and Zhao, Yi},
  journal = {J. Phys. Chem. Lett.},
  volume = {17},
  issue = {25},
  pages = {6398--6404},
  year = {2026},
  doi = {10.1021/acs.jpclett.6c00947}
}

@article{MorillasRozas2026,
  title = {The Transfer Tensor Method: An Analytical Study Case},
  author = {Morillas-Rozas, Marcel and L{\'o}pez-Garc{\'i}a, Alberto and Reina Rivero, Gonzalo and Cao, Jianshu and Cerrillo, Javier},
  journal = {J. Chem. Phys.},
  volume = {165},
  pages = {054108},
  year = {2026}
}

@article{Noh2010,
  title = {Analytic solutions of the {B}loch equations for a driven two-level atom with relaxation},
  author = {Noh, Heung-Ryoul and Jhe, Wonho},
  journal = {Opt. Commun.},
  volume = {283},
  issue = {11},
  pages = {2353--2355},
  year = {2010},
  doi = {10.1016/j.optcom.2010.01.077}
}

@article{Shin2024,
  title = {Room-temperature optical cycling molecules in liquid cells},
  author = {Shin, D. and Alexandrova, A. N. and Diaconescu, P. L. and Campbell, W. C. and Caram, J. R.},
  journal = {Science},
  volume = {385},
  issue = {6709},
  pages = {651--656},
  year = {2024},
  doi = {10.1126/science.adn3230}
}

@article{Mickiewicz2026,
  title = {Benchmarking {F}loquet Master Equations for Periodically Driven Open Quantum Systems},
  author = {Mickiewicz, Konrad and Link, Valentin and Strunz, Walter T.},
  journal = {arXiv preprint arXiv:2606.06341},
  year = {2026}
}

@article{Cao2023,
  title = {Driving induced coherent quantum energy transport},
  author = {Cao, Xiufeng and Wang, Chen and He, Dahai},
  journal = {Phys. Rev. B},
  volume = {108},
  pages = {245401},
  year = {2023},
  doi = {10.1103/PhysRevB.108.245401}
}

@article{Cao2026,
  title = {Steady-state quantum coherence in driven open quantum system: An optimal transformation analysis},
  author = {Cao, Xiufeng and Wang, Chen and Engelhardt, Georg and Zheng, Hang and He, Dahai},
  journal = {J. Chem. Phys.},
  volume = {164},
  pages = {164307},
  year = {2026},
  doi = {10.1063/5.0325263}
}

@article{PollockModi2018,
  title = {Tomographically reconstructed master equations for any open quantum dynamics},
  author = {Pollock, Felix A. and Modi, Kavan},
  journal = {Quantum},
  volume = {2},
  pages = {76},
  year = {2018},
  doi = {10.22331/q-2018-07-11-76}
}

@article{Jorgensen2020,
  title = {Discrete memory kernel for multitime correlations in non-{M}arkovian quantum processes},
  author = {J{\o}rgensen, Mathias R. and Pollock, Felix A.},
  journal = {Phys. Rev. A},
  volume = {102},
  pages = {052206},
  year = {2020},
  doi = {10.1103/PhysRevA.102.052206}
}

@article{Gherardini2022,
  title = {Transfer-tensor description of memory effects in open-system dynamics and multi-time statistics},
  author = {Gherardini, Stefano and Smirne, Andrea and Huelga, Susana F. and Caruso, Filippo},
  journal = {Quantum Sci. Technol.},
  volume = {7},
  pages = {025005},
  year = {2022},
  doi = {10.1088/2058-9565/ac4422}
}

@article{gelzinis2017applicability,
  title={Applicability of transfer tensor method for open quantum system dynamics},
  author={Gelzinis, Andrius and Rybakovas, Edvardas and Valkunas, Leonas},
  journal={The Journal of Chemical Physics},
  volume={147},
  number={23},
  pages={234108},
  year={2017},
  doi={10.1063/1.5009086},
  publisher={AIP Publishing}
}

@article{chen2020non,
  title={Non-{M}arkovian noise characterization with the transfer tensor method},
  author={Chen, Yu-Qin and Ma, Kai-Li and Zheng, Yi-Cong and Allcock, Jonathan and Zhang, Shengyu and Hsieh, Chang-Yu},
  journal={Physical Review Applied},
  volume={13},
  number={3},
  pages={034045},
  year={2020},
  publisher={APS}
}

@article{zeng2026estimating,
  title={Estimating memory time within the frameworks of generalized quantum master equation and transfer tensor methods},
  author={Zeng, Hao and Sun, Xiang},
  journal={The Journal of Chemical Physics},
  volume={164},
  number={22},
  pages={224117},
  year={2026},
  doi={10.1063/5.0325277},
  publisher={AIP Publishing}
}

@article{jirari2026analytical,
  title={Analytical insights into the spin--boson model and optimal control via the {R}edfield master equation},
  author={Jirari, H and Elamrani, A and Rabitz, H},
  journal={The European Physical Journal B},
  volume={99},
  number={5},
  pages={62},
  year={2026},
  publisher={Springer}
}

@article{settimo2026synchronization,
  title={Synchronization effects in a periodically driven two-level system},
  author={Settimo, Federico and Vacchini, Bassano},
  journal={Physical Review A},
  volume={113},
  number={2},
  pages={022213},
  year={2026},
  publisher={APS}
}

@article{Mickiewicz2025,
  title = {Exact {F}loquet dynamics of strongly damped driven quantum systems},
  author = {Mickiewicz, Konrad and Link, Valentin and Strunz, Walter T.},
  journal = {Phys. Rev. Lett.},
  volume = {136},
  pages = {200201},
  year = {2026},
  doi = {10.1103/5z1m-122d}
}

@article{Breuer2018,
  title = {Mixing-induced quantum non-{M}arkovianity and information flow},
  author = {Breuer, Heinz-Peter and Amato, Giulio and Vacchini, Bassano},
  journal = {New J. Phys.},
  volume = {20},
  pages = {043007},
  year = {2018},
  note = {arXiv:1712.06719}
}

@article{Lorenzo2017,
  title = {Non-{M}arkovian dynamics from band edge effects and static disorder},
  author = {Lorenzo, Salvatore and Ciccarello, Francesco and Palma, G. Massimo},
  journal = {arXiv preprint arXiv:1712.06670},
  year = {2017}
}

@article{Gangwar2026,
  title = {Genuine and Non-Genuine Quantum Non-{M}arkovianity: A Unified Information-Theoretic Review},
  author = {Gangwar, Rajeev and Sen, Ujjwal},
  journal = {arXiv preprint arXiv:2603.28277},
  year = {2026}
}

@article{Sengupta2026,
  title = {Can the system dynamics remain structurally non-{M}arkovian while its experimentally accessible memory signatures disappear?},
  author = {Sengupta, Sayan and Sharma, Anish and Biswas, George and Biswas, Anindya},
  journal = {arXiv preprint arXiv:2610.18582},
  year = {2026}
}

@article{Santra2024,
  title = {Disorder-averaged Qudit Dynamics},
  author = {Santra, Gopal Chandra and Hauke, Philipp},
  journal = {arXiv preprint arXiv:2412.17519},
  year = {2024}
}

@article{Chaudhary2026,
  title = {Reinterpreting Memory Effects in Nonequilibrium Systems: From Temporal Dynamics to Steady-State Signatures via {NEGF}},
  author = {Chaudhary, Pragya},
  journal = {arXiv preprint arXiv:2605.28993},
  year = {2026}
}

@article{Liu2026,
  title = {Coarse-Grained Dynamics with Spatial Disorder and Non-{M}arkovian Memory},
  author = {Liu, Chuyi and Guan, Yifeng and Li, Jingyuan and Su, Mao},
  journal = {arXiv preprint arXiv:2604.16825},
  year = {2026}
}

@article{Engelhardt2023,
  title = {Polariton Localization and Dispersion Properties of Disordered Quantum Emitters in Multimode Microcavities},
  author = {Engelhardt, Georg and Cao, Jianshu},
  journal = {Phys. Rev. Lett.},
  volume = {130},
  pages = {213602},
  year = {2023},
  doi = {10.1103/PhysRevLett.130.213602}
}

@article{Aroeira2025,
  title = {Static disorder-induced renormalization of polariton group velocity},
  author = {Aroeira, Gustavo J. R. and Ribeiro, Raphael F.},
  journal = {arXiv preprint arXiv:2507.00918},
  year = {2025}
}

@article{Holthaus1995,
  title = {ac-Field-Controlled {A}nderson Localization in Disordered Semiconductor Superlattices},
  author = {Holthaus, Martin and Ristow, Gerald H. and Hone, Daniel W.},
  journal = {Phys. Rev. Lett.},
  volume = {75},
  pages = {3914},
  year = {1995},
  doi = {10.1103/PhysRevLett.75.3914}
}

@article{Holthaus1996,
  title = {Localization effects in ac-driven tight-binding lattices},
  author = {Holthaus, Martin and Hone, Daniel W.},
  journal = {Philos. Mag. B},
  volume = {74},
  pages = {105},
  year = {1996},
  doi = {10.1080/01418639608240331}
}

@article{Martinez2006,
  title = {Localization properties of driven disordered one-dimensional systems},
  author = {Martinez, Dario F. and Molina, Rafael A.},
  journal = {Eur. Phys. J. B},
  volume = {52},
  pages = {281--290},
  year = {2006},
  doi = {10.1140/epjb/e2006-00293-7}
}

@article{Zhang2006,
  title = {{AC}-field-controlled localization-delocalization transition in one dimensional disordered system},
  author = {Zhang, Wei and Ulloa, Sergio E.},
  journal = {arXiv preprint cond-mat/0603720},
  year = {2006}
}

@article{Dotti2025,
  title = {Measuring a localization phase diagram controlled by the interplay of disorder and driving},
  author = {Dotti, Peter and Bai, Yifei and Shimasaki, Toshihiko and Dardia, Anna R. and Weld, David M.},
  journal = {Phys. Rev. Res.},
  volume = {7},
  pages = {L022026},
  year = {2025},
  doi = {10.1103/PhysRevResearch.7.L022026}
}

@article{Schindler2025,
  title = {Floquet engineering with spatially nonuniform driving fields},
  author = {Schindler, Stella T. and Herzig Sheinfux, Hanan},
  journal = {Photonics Res.},
  volume = {13},
  pages = {905},
  year = {2025}
}

@article{Tiwari2024,
  title = {Quantum transport under oscillatory drive with disordered amplitude},
  author = {Tiwari, Vatsana and Dattagupta, Sushanta and Bhakuni, Devendra Singh and Sharma, Auditya},
  journal = {arXiv preprint arXiv:2408.12653},
  year = {2024}
}

@article{Satyajit2026,
  title = {Eigenframe Synchronization in Disordered Driven Quantum Ensembles},
  author = {Satyajit, Saipriya and Yin, Zechuan and Oon, Jner Tzern and Everaert, Katrijn and Bhalerao, Smriti and Blanchard, John W. and Jarzynski, Christopher and Walsworth, Ronald L.},
  journal = {arXiv preprint arXiv:2608.17026},
  year = {2026}
}

@article{Ducatez2017,
  title = {Anderson Localization for Periodically Driven Systems},
  author = {Ducatez, Raphael and Huveneers, Fran{\c{c}}ois},
  journal = {Ann. Henri Poincar{\'e}},
  volume = {18},
  pages = {2415--2446},
  year = {2017},
  doi = {10.1007/s00023-017-0574-1}
}

@article{Agarwal2017,
  title = {Localization and transport in a strongly driven Anderson insulator},
  author = {Agarwal, Kartiek and Ganeshan, Sriram and Bhatt, R. N.},
  journal = {Phys. Rev. B},
  volume = {96},
  pages = {014201},
  year = {2017},
  doi = {10.1103/PhysRevB.96.014201}
}

@article{Wauters2019,
  title = {Localization, Topology, and Quantized Transport in Disordered {F}loquet Systems},
  author = {Wauters, Matteo M. and Russomanno, Angelo and Citro, Roberta and Santoro, Giuseppe E. and Privitera, Lorenzo},
  journal = {Phys. Rev. Lett.},
  volume = {123},
  pages = {266601},
  year = {2019},
  doi = {10.1103/PhysRevLett.123.266601}
}

@misc{FessahayeZenodo2026,
  author    = {Fessahaye, Ferdos and Dutta, Rajesh and Anderson, Michelle and Chuang, Chern},
  title     = {Data and code for ``{Transfer tensor description of driven inhomogeneous quantum systems}''},
  year      = {2026},
  note      = {{Zenodo}, version v1.0, DOI: 10.5281/zenodo.24492393},
  doi       = {10.5281/zenodo.24492393}
}

\end{document}